\documentclass[twocolumn,twocolappendix]{aastex631}
\usepackage{CJK}
\usepackage{bm}
\usepackage{appendix}
\usepackage{amsmath,amssymb}
\usepackage{natbib}

\newcommand{\code}[1]{\texttt{#1}}

\newcommand{\logM}{\log (M_{\star}/M_{\sun})}

\defcitealias{Haslbauer:2022}{H22}

\shorttitle{}
\shortauthors{Xu et al.}

\graphicspath{{./}{figures/}}

\begin{document}
\begin{CJK*}{UTF8}{gbsn}

\correspondingauthor{Hua Gao}

\title{IllustrisTNG in the HSC-SSP: No Shortage of Thin Disk Galaxies in TNG50}

\author[0000-0003-2154-2098]{Dewang Xu (徐德望)}
\affiliation{Department of Astronomy, School of Physics, Peking University, Beijing 100871, People's Republic of China}
\affiliation{Kavli Institute for Astronomy and Astrophysics, Peking University, Beijing 100871, People's Republic of China}

\author[0000-0003-1015-5367]{Hua Gao (高桦)}
\affiliation{Institute for Astronomy, University of Hawaii, 2680 Woodlawn Drive, Honolulu, HI 96822, USA}
\email{hgao.astro@gmail.com}

\author[0000-0003-4758-4501]{Connor Bottrell}
\affiliation{International Centre for Radio Astronomy Research, University of Western Australia, 35 Stirling Hwy, Crawley, WA 6009, Australia}

\author[0000-0002-4176-9145]{Hassen M. Yesuf}
\affiliation{Key Laboratory for Research in Galaxies and Cosmology, Shanghai Astronomical Observatory, Chinese Academy of Sciences, 80 Nandan Road, Shanghai 200030, China}
\affiliation{Kavli Institute for the Physics and Mathematics of the Universe (WPI), The University of Tokyo Institutes for Advanced Study (UTIAS), The University of Tokyo, 5-1-5 Kashiwanoha, Kashiwa-shi, Chiba 277-8583, Japan}
\affiliation{Kavli Institute for Astronomy and Astrophysics, Peking University, Beijing 100871, People's Republic of China}

\author[0000-0001-9879-4926]{Jingjing Shi}
\affiliation{Kavli Institute for the Physics and Mathematics of the Universe (WPI), The University of Tokyo Institutes for Advanced Study (UTIAS), The University of Tokyo, 5-1-5 Kashiwanoha, Kashiwa-shi, Chiba 277-8583, Japan}
\affiliation{Center for Data-Driven Discovery (CD3), Kavli IPMU (WPI), UTIAS, The University of Tokyo, Kashiwa, Chiba 277-8583, Japan}

\begin{abstract}
We perform a thorough analysis of the projected shapes of nearby galaxies in both observations and cosmological simulations. We implement a forward-modeling approach to overcome the limitations in previous studies, which hinder accurate comparisons between observations and simulations. We measure axis ratios of $z=0$ (snapshot 99) TNG50 galaxies from their synthetic Hyper Suprime-Cam Subaru Strategic Program (HSC-SSP) images and compare them with those obtained from real HSC-SSP images of a matched galaxy sample. Remarkably, the comparison shows excellent agreement between the observations and the TNG50 simulation, challenging previous claims that $\Lambda$CDM models underproduced the abundance of thin galaxies. Specifically, for galaxies with stellar masses $10\leq \logM \leq 11.5$, we find $\lesssim 0.1\sigma$ tensions between the observations and the simulation, a stark contrast to the previously reported $\gtrsim 10\sigma$ tensions. We reveal that low-mass galaxies ($M_{\star}\lesssim 10^{9.5}\,M_{\odot}$) in TNG50 are thicker than their observed counterparts in HSC-SSP and attribute this to the spurious dynamical heating effects that artificially puff up galaxies. We also find that, despite the overall broad agreement, TNG50 galaxies are more concentrated than the HSC-SSP ones at the low- and high-mass end of the stellar mass range of $9.0\leq \logM \leq 11.2$ and are less concentrated at intermediate stellar masses. But we argue that the higher concentrations of the low-mass TNG50 galaxies are not likely the cause of their thicker/rounder appearances. Our study underscores the critical importance of conducting mock observations of simulations and applying consistent measurement methodologies to facilitate proper comparison with observations.
\end{abstract}

\keywords{Hydrodynamical simulations (767) --- Galaxy evolution (594) --- Galaxy kinematics(602) --- Galaxy photometry (611) --- Galaxy structure (622)}

\section{Introduction}\label{sec:intro}

The diverse morphological structures of nearby galaxies along the Hubble sequence \citep{Hubble:1926, 1936rene+Hubble, 1959HDP+de_Vaucouleurs}, ranging from spheroidal to flattened shapes, are imprints of distinct formation and evolutionary pathways of galaxies. The axis ratios of galaxies ($q$) quantify the projected shapes of their 3D galactic structures on the sky plane. Notably, the axis ratio encapsulates the combined contribution from all substructures within a galaxy. The distribution of axis ratios has long been observed to differ among various galaxy populations \citep[e.g.,][]{Sandage:1970, Binggeli:1980, Ryden:1994, Im:1995}, thereby providing useful probes of their evolutionary history. Analyzing distinct axis ratio distributions enables inference of the intrinsic 3D shapes of these galaxies, under various assumptions of ellipsoids \citep[e.g.,][]{Sandage:1970, Lambas:1992, Padilla:2008, Law:2012, chang:2013b, vanderWel:2014, Zhang:2022threeDintrin, Pandya+2024banana}. A simpler metric of the intrinsic shape of galaxies, the edge-on thickness ($q_0$), can be inferred from their axis ratio distributions \citep[e.g.,][]{Sanchez-Janssen:2010}. As expected, $q_0$ varies with other physical properties of galaxies, such as Hubble types \citep[e.g.,][]{Sandage:1970, Fouque1990}, colors \citep[e.g.,][]{Padilla:2008, Masters2010, Rodriguez:2013}, and stellar masses \citep[e.g.,][]{Sanchez-Janssen:2010, Holden:2012}. Interestingly, \citet{Sanchez-Janssen:2010} identified a U-shaped dependence of $q_0$ on stellar mass.
 
Comparing observations with cosmological predictions is a critical test of galaxy formation and evolution theories and also helps the interpretation of the observations. In addition to reproducing the fundamental scaling relations of galaxies, examining the detailed galaxy morphologies adds to the comprehensive measure of the accuracy and success of the models. \citet{2009ApJ+Weinzirl} found that bulges in semi-analytic $\Lambda$CDM models are more prominent than the observed ones. Moreover, other studies claimed that there were an excessive number of bulgeless galaxies and galaxies with pseudo bulges compared to what they expected from hierarchical clustering \citep{2008ASPC+Kormendy, Kormendy:2010}. However, it is worth noting that statistics of bulgeless galaxies and pseudo bulges are sensitive to the measurement techniques and selection criteria \citep{Gao+2017, Gao+2018, Gao:2019, Gao+2020, Gao+2022}. Recent developments in hydrodynamical cosmological simulations, such as Illustris \citep{Vogelsberger:2013, Torrey:2014}, IllustrisTNG \citep[hereafter TNG;][]{Weinberger:2017, Pillepich:2018a}, and EAGLE \citep{Crain:2015, Schaye:2015}, have addressed the issue of angular momentum loss and produced more realistic looking galaxies \citep[e.g.,][]{Vogelsberger:2014}. Still, tensions remain between the theories and observations, probably due to limitations in numerical resolution and physical models. Studies of both Illustris and EAGLE reported a significant deficit of bulge-dominated galaxies in the simulations \citep{Bottrell:2017a, Bottrell:2017b, deGraaff:2022}. On the contrary, \citeauthor{Haslbauer:2022} (\citeyear{Haslbauer:2022}; hereafter \citetalias{Haslbauer:2022}) reported a lack of thin galaxies in Illustris, TNG, and EAGLE, despite that their median axis ratios as a function of stellar mass agree well with observations \citep{Rodriguez-Gomez:2019}. Adding to this, \citet{vandeSande:2019} showed a lack of highly flattened galaxies in various simulations including EAGLE, HYDRANGEA \citep{Bahe:2017}, and HORIZON-AGN \citep{Dubois2014}, except for Magneticum Pathfinder \citep{Hirschmann2014, Teklu2015}. The consistencies and conflicts demand more rigorous and self-consistent comparisons between observations and simulations to better understand the origins of such discrepancies.

Some studies suffer from inconsistent treatment of the observational data and the simulations. Simulations offer 6D phase space information of the basic elements (e.g., stellar/dark matter particles and gas cells), while in observations only projected distribution and line-of-sight velocities of baryons are directly available. Because of the difference in availability of information and formats of data, authors often apply different sample selection criteria and measurement techniques to both sides, which helps to maximize the outcome from both but might have resulted in inaccurate conclusions. For example, \citetalias{Haslbauer:2022} applied different methods to measure galaxy shapes from simulations and observations, which may have significant impacts on the fairness of the comparison. 
Meanwhile, significant effort has been devoted to ensuring the measurement methods are consistent; however, the consistency is often compromised by lack of realistic mock observation of the simulations \citep[e.g.,][]{vandeSande:2019}. Moreover, \citet{deGraaff:2022} showed that using synthetic data of full realism helps alleviate the tension between the simulated and observed galaxy mass--size relation. We are therefore motivated to carry out a self-consistent comparison of galaxy shapes using synthetic images of simulated galaxies with full realism and aim to permanently resolve the controversy.

Synthetic data have long been used to bridge the gap between simulations and observations for the purpose of comparison \citep[e.g.,][]{Bottrell:2017a, Bottrell:2017b, Rodriguez-Gomez:2019, Eisert2024contrastive}. In particular, radiative transfer post-processing of randomly oriented simulated galaxies to create mock images offers several advantages over previous studies: (1) we do not need to make assumptions of the 3D structure of the simulated galaxies for projection, (2) they automatically resolve the inconsistencies concerning mass-to-light ratios and instrumental effects, and (3) they allow application of the same measurement techniques to both sides. We will use a set of synthetic Hyper Suprime-Cam Subaru Strategic Program \citep[HSC-SSP;][]{Aihara:2018b} images of TNG50 galaxies with $\logM \geq 9$ in the redshift range $0 \leq z \leq 0.7$ created by \citet{Bottrell:2023}. Data and sample selection are described in Section~\ref{sec:data}.
To mitigate systematics, we apply the same method to measure axis ratios of galaxies from real HSC-SSP images and the synthetic HSC-SSP images of TNG50, as outlined in Section~\ref{sec:method}. Results and discussions are given in Sections~\ref{sec:results} and \ref{sec:discussions}. We summarize the paper in Section~\ref{sec:summary}. We adopt a cosmology based on Planck 2015 \citep{PlanckCollaboration:2016} with $H_0 = 100h$\,km s$^{-1}$\,Mpc$^{-1}$ with $h=0.6774$, $\Omega_m = 0.308$, and $\Omega_\Lambda = 0.692$.

\section{Data and Sample} \label{sec:data}
\subsection{The Hyper Suprime-Cam Subaru Strategic Program Survey}

The HSC-SSP \citep{Aihara:2018b} is an optical imaging survey that uses the HSC \citep{Miyazaki2012, Miyazaki2018} of the $8.2$-meter Subaru Telescope. HSC-SSP has released three public data releases \citep[PDRs;][]{Aihara:2018b, Aihara:2019, Aihara:2022} and covers about $1000\,\mathrm{deg^2}$ in $grizy$ bands in the \texttt{WIDE} layer. One can reliably extract the surface brightness profiles of galaxies down to $28.5\,\mathrm{mag\, arcsec^{-2}}$ in $r$ band and $i$ band using PDR2 \texttt{WIDE} images \citep{Huang:2018individualHalo, Li:2022}. The best spatial resolution is in the $i$-band, with a median seeing of $0\farcs61$ for the PDR3 \texttt{WIDE} layer. The HSC has a pixel scale of $0\farcs168$.

The HSC-SSP photometric catalog\footnote{The photometric catalogs are described in \url{https://hsc-release.mtk.nao.ac.jp/schema/}.} has incorporated publicly available spectroscopic redshifts (spec-$z$) from various sources such as the Galaxy And Mass Assembly \citep[GAMA;][]{Driver:2009, Driver:2011} DR2 \citep{Liske:2015} and the Sloan Digital Sky Survey \citep[SDSS;][]{SDSSCollaboration:2000} DR16 \citep{Ahumada:2020}. A homogenized flag in the photometric catalog indicates which spec-$z$ has been adopted out of the various available sources, opting for the spec-$z$ with the smallest error \citep{Aihara:2018b}. We select all primary objects with known spec-$z$ by querying the photometric catalog with $\mathtt{isprimary = True}$ and $\mathtt{specz\_redshift}>0$. Galaxies are selected from the photometric catalog based on $\mathtt{extendedness=1}$. Furthermore, we select full-depth galaxy images of superior quality based on the following flags: $\mathtt{g\_inputcount\_value\geq4}$, and $\mathtt{(i\&y)\_inputcount\_value\geq5}$, \texttt{NOT pixelflags\_edge}, \texttt{NOT pixelflags\_offimage}, \texttt{NOT mask\_brightstar\_halo}, \texttt{NOT mask\_brightstar\_blooming}.
We cross-match the HSC-SSP photometric catalog with the stellar mass and size measurements from GAMA \citep{Taylor:2011} and SDSS \citep{Ahn:2014}. The stellar mass for a galaxy whose spec-$z$ from either SDSS or GAMA, is obtained from the measurements of the corresponding source. \citet{Taylor:2011} estimate the stellar masses using the \citet{Chabrier:2003} initial mass function and \citet{BruzualCharlot2003} stellar population synthesis models, while \citet{Ahn:2014} use the \citet{Kroupa:2001} initial mass function and adopts the Flexible Stellar Population Synthesis code \citep[FSPS;][]{Conroy:2009}. We confirm a systematic offset between the two measurements of stellar mass, as reported in \citet{Kado-Fong:2020SFR, Kado-Fong:2020Shape}. The SDSS stellar masses are systematically higher than the GAMA ones, with a mean offset of $0.17$\,dex and a scatter of $0.26$\,dex estimated from the overlapped galaxies. To align the measurements and to project consistency with the generation of the synthetic images (Section~\ref{sec:syn-img}), we subtract the offset of $0.17$\,dex from the SDSS stellar masses. To compare with the TNG50 simulation, we select our main sample of 4657 galaxies within a redshift range of $0.005 \leq z \leq 0.05$ and with a stellar mass range of $9 \leq \logM \leq 11.5$. Furthermore, to gain a comprehensive view of the axis ratio distribution including dwarf galaxies, we also include galaxies with smaller stellar masses of $8.5\leq \logM <9$.

In this work, we use the $g$, $i$, and $y$-band images from the HSC-SSP PDR3 \texttt{WIDE} layer. We extract cutout images with a size of $15\, r_{90} \times 15\, r_{90}$, where $r_{90}$ is the radius within which the integrated flux constitutes $90\%$ of the total galaxy flux. The value of $r_{90}$ is obtained from the same source as the spec-$z$ (either SDSS or GAMA). We also retrieve point spread function (PSF) images at the center of each individual galaxy. We use the command-line SQL tool\footnote{\url{https://hsc-gitlab.mtk.nao.ac.jp/ssp-software/data-access-tools/-/tree/master/pdr3}} to download all image products. 

\subsection{Synthetic HSC-SSP Images of TNG50}\label{sec:syn-img}

The TNG project\footnote{\url{http://www.illustris-project.org/}} is a suite of magneto-hydrodynamic cosmological simulations of the formation and evolution of various elements, including gas, stars, dark matter, supermassive black holes, and magnetic fields from $z=127$ to $z=0$ \citep{Weinberger:2017, Pillepich:2018a, Nelson2019a, Nelson2019outflow, Pillepich2019}. The TNG suite consists of three simulations, respectively referred to as TNG50, TNG100, and TNG300. They run at $V_\mathrm{box}$ volumes of $51.7^3$, $106.5^3$, and $302.6^3$ cMpc$^3$, respectively. Two commonly used resolution indicators are (1) the Plummer-equivalent gravitational softening length for collisionless particles, $\epsilon_{\star, \mathrm{dm}} = (0.29, 0.74, 1.48)$ ckpc, and (2) the average size of star-forming gas cells, $\bar{r}_\mathrm{cell,sf}= (0.138, 0.355, 0.715)$ ckpc. 

We use the synthetic HSC-SSP \texttt{WIDE} images generated by \cite{Bottrell:2023} from the TNG50 simulation. Our work focuses on simulated galaxies in snapshot 99 ($z=0$) mock observed at $z=0.05$. The photo-realistic but noise-free (idealized) synthetic images in HSC $grizy$ bands were first generated using \code{SKIRT} Monte Carlo radiative transfer \citep[][\code{SKIRT} version 9]{Camps&Baes:2020}. Using a bespoke version of RealSim\footnote{\url{https://github.com/cbottrell/RealSim}} for the HSC-SSP \citep{Bottrell2019realsim}, the idealized images made by \code{SKIRT} were then convolved with reconstructed HSC-SSP seeing kernels and statistically injected into final-depth HSC-SSP cutouts so that these synthetic images have consistent properties with observed HSC-SSP \code{WIDE} layer images of real galaxies. The TNG50 snapshot 99 includes $3092$ galaxies with $9 \leq \logM \leq 11.5$, and their synthetic HSC-SSP galaxy images are generated for each TNG50 galaxy at four different viewing angles, resulting in a final sample of $12368$ synthetic images. We refer readers to \citet{Bottrell:2023} for a detailed description of the procedure.

\begin{figure}
     \centering
	\includegraphics[width=1\linewidth]{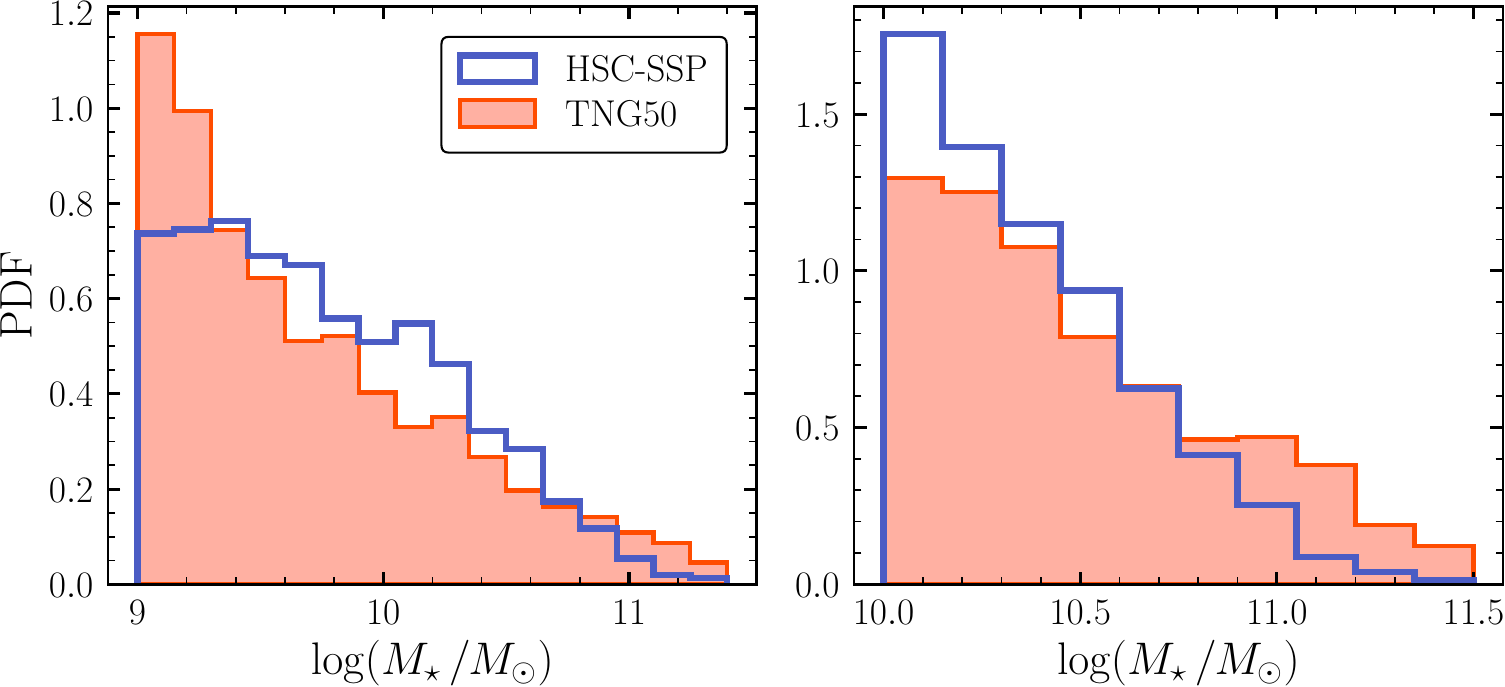}
     \caption{Stellar mass distributions of the HSC-SSP galaxies (open blue) and the TNG50 galaxies (filled red). Left: The main sample of TNG50 galaxies and HSC-SSP galaxies in the stellar mass range of $9 \leq \logM \leq 11.5$. Right: high-mass samples with $10 \leq \logM \leq 11.5$.}
     \label{fig:mass_distribution} 
\end{figure}

Figure~\ref{fig:mass_distribution} compares the stellar mass distributions of the observed HSC-SSP sample ($0.005 \leq z \leq 0.05$) and the TNG50 sample (snapshot 99 mock observed at $z=0.05$).

\section{Methodology}\label{sec:method}
\subsection{Object Masks}\label{sec:mask}

We use the \texttt{sep} package\footnote{\url{https://sep.readthedocs.io/en/v1.1.x}} \citep{Barbary:2016}, a \code{Python} implementation of \texttt{SourceExtractor}, to detect and deblend objects using the \texttt{cold + hot} approach \citep[e.g.,][]{Rix:2004}. We run \texttt{sep} with two sets of parameters. The \texttt{hot mode} detects bright objects and ensures a complete segmentation of the target galaxy by using a $5 \sigma$ detection threshold and $\mathtt{deblend\_cont}=0.1$. The \texttt{cold mode} detects faint galaxies and stars by using a $1 \sigma$ detection threshold and $\mathtt{deblend\_cont} = 0.001$. We first obtain the position and size information of the target from the \texttt{hot mode}. Within $3$ times the Kron radius of the target, we remove all the objects detected by the \texttt{cold mode}, except for those that cover more than $80\%$ of the area of the target galaxy in the \texttt{hot mode}. Then, we remove the target from both the \texttt{cold mode} and the \texttt{hot mode}, and combine the remaining objects to create masks that we use for further analysis.
We dilate the object masks based on the distance of each object from the galaxy center. For objects located more than $6\, r_{50}$ away from the target center, where $r_{50}$ is the radius within which $50\%$ of the total flux is enclosed, we apply a Gaussian filter with a sigma of $3$ times the median FWHM of PSFs in each band. For those objects located within $6\, r_{50}$ of the center, we use an elliptical aperture mask whose semi-major (minor) axis lengths are the second order moments of the object's flux along the major (minor) axis given by \code{sep}. We combine the mask with the bad pixel mask (either `NO\_DATA' or `SAT' mask planes) provided by HSC-SSP PDR3 to form the final masks.

The above analysis is performed for both real and synthetic HSC-SSP images. Moreover, the final masks for the TNG50 synthetic images also include the detected source footprints as part of the original HSC-SSP image products, to err on the safe side despite that they should mostly overlap with the \code{sep} masks. This extra step does not apply to the real images because the footprints include the galaxy of interest and are not deblended.

\subsection{Measurements of Global Axis Ratios}\label{sec:q_meas}

\code{AutoProf}\footnote{\url{https://autoprof.readthedocs.io/en/latest}} is an automatic nonparametric tool for fitting isophote and extracting surface brightness profiles of extended sources \citep{Stone:2021}. The code has several modular steps that can be customized by the user to form a desired pipeline of data processing. For our application, the full feature of \code{AutoProf} is more than enough. We only use the initialization step to measure the global axis ratio ($q$) and position angle (PA) of the galaxies. We supply the masks prepared in Section~\ref{sec:mask} and PSF FWHMs obtained via 2D Gaussian fits as input for \code{AutoProf}. Moreover, we need quality flags produced by the checking fit step to help discard poor fits. Below, we provide a brief overview of these two steps. For further details, please refer to Sections~2.4~and~2.7 of \citet{Stone:2021}.

The initialization is done in two steps. First, fast Fourier transform (FFT) is performed over fluxes along circular paths concentric with the galaxy, and the phases are determined from the second Fourier coefficient $\mathcal{F}_2$. The global PA is the average direction of the phases for the five outermost circular paths whose average fluxes are still above the background level. Second, with the global PA being fixed, the global axis ratio is obtained by minimizing $\left|\mathcal{F}_{2} /\left(\tilde{\mu}+\sigma_{b}\right)\right|$, where $\tilde{\mu}$ is the median flux, $\sigma_{b}$ is the background noise, and $\tilde{\mu}+\sigma_{b}$ is approximately equal to $\mathcal{F}_0$.

The checking fit step performs four checks on the quality of the fits by examining the flux distribution along the best-fit isophotes: (1) and (2) the magnitudes of the first and the second Fourier mode, respectively; (3) the variability of fluxes; and (4) the variability of fluxes compared with that along the initialized isophotes. The first check failing due to significant power in the first Fourier mode often indicates a failed center determination or lopsidedness in galaxy flux distribution, and the second suggests that the optimization was not achieved or the presence of elongated structures (e.g., bars). The third check compares the interquartile range of the fluxes to their median and fails when a minimum fraction of the isophotes have such high variabilities. The fourth check is similar to the third, only that the variability is contrasted against that obtained from the initialized isophotes. The thresholds for triggering the failures were chosen by \citet{Stone:2021} to catch visually identified cases.


Upon visual inspection of a random sample of 500 galaxies, we find that the fits that have passed at least three checks are of good quality. Hence, galaxies that failed more than one check are removed from the sample, which constitutes a small fraction of our sample (e.g., $\sim 3\%$ for HSC-SSP $i$-band observations). Additionally, we implement a resolution cut to ensure sufficient spatial resolution for the initialized isophotes where $q$ was measured. Specifically, we required a semi-minor axis length of the initialized isophote larger than $1.5$ times the PSF FWHM, i.e., $R \times q \geq 1.5 \times \mathrm{FWHM}$, where $R$ is the semi-major axis length of the initialized isophote. The resolution cut has minimal impact on the sample, even in the case of the $y$-band images, which exhibit the worst seeing statistics (see Appendix~\ref{appendix:rescut} for details). 

\subsection{Measurements of Structural Parameters}

In this section, we describe how we measure the concentration index from the (synthetic) HSC-SSP images and how the spheroidal-to-total ratio (S/T) is derived from stellar kinematics in the simulation.

We adopt the curve-of-growth method to derive the concentration index for both the HSC-SSP galaxies and the TNG50 galaxies. We use \code{AutoProf} to extract 1D, azimuthally averaged surface brightness profiles of the galaxies. Subsequently, based on the growth curve of their flux truncated at the limiting surface brightness, we derive $r_{20}$ and $r_{80}$, where $r_{20}$ ($r_{80}$) represents the radius within which the integrated flux constitutes $20\%$ ($80\%$) of the total galaxy flux. We define the concentration index as $C_{82} = 5 \times \log{(r_{80}/r_{20})}$. We exclude galaxies with $r_{20}< 1.5 \times {\rm FWHM}$ to ensure sufficient spatial resolution for accurate measurements. The index characterizes the degree of flux concentration of a galaxy and is often used to separate bulge-dominated and disk-dominated galaxies \citep[e.g.,][]{Conselice:2003, Conselice:2014}.

S/T of a galaxy quantifies the fraction of its stellar mass in the spheroidal (kinematically hot) components. We follow the definition of \citet{Tacchella:2019} that stars of spheroidal components have $\epsilon \leq 0.7$, where $\epsilon$ is the circularity of stellar particles. \citet{Genel:2015} provided the fractional mass of stars with $\epsilon > 0.7$ computed within ten times the stellar half-mass radii of TNG galaxies as \code{CircAbove07Frac}\footnote{\url{https://www.tng-project.org/data/docs/specifications/\#sec5c}}. So, we simply derive $S/T=1-\mathtt{CircAbove07Frac}$.

\section{Results}\label{sec:results}






\subsection{Comparison of Overall Axis Ratio Distributions between HSC-SSP and TNG50} \label{sec:res_qdis}

We aim to conduct a fair comparison of axis ratios between the HSC-SSP galaxies and the TNG50 galaxies. We apply the same method defined in Section~\ref{sec:q_meas} to both the synthetic HSC-SSP images of TNG50 galaxies and the real HSC-SSP images, in order to obtain comparable measurements of galaxy shapes. 
The comparison is performed within the stellar mass range of $10 \leq \logM \leq 11.5$. The upper limit of the stellar mass range was set to ensure nonzero presence of galaxies in the highest stellar mass bins. 
We quantify the difference between the observation and the simulation using chi-square ($\chi^2$) statistics and the associated statistical significance. The formulation of $\chi^2$ that describes the difference between the axis ratio distributions of the observed and simulated galaxies is given below. 
\begin{equation}\label{eq:chisq_weight}
    \chi^{2}=\sum_{i=1}^{N_{\mathrm{bins},q}} \frac{(N^{\prime}_{\mathrm{obs}, i} - N_{\mathrm{sim},i})^{2}}{\sigma_{\mathrm{obs}, i}^{2}+\sigma_{\mathrm{sim}, i}^{2}},
\end{equation}
\begin{equation}\label{eq:sigma_obs}
    \sigma_{{\rm obs},i} = \sqrt{\sum_{j=1}^{N_{{\rm bins},M_\star}}{w_{{\rm obs},j}^2 \max (N_{\mathrm{obs},i,j}}, 1) },
\end{equation}
\begin{equation}\label{eq:sigma_mod}
    \sigma_{{\rm sim},i} = \sqrt{\max (N_{{\rm sim},i}, 1)},
\end{equation}
where $N_{\mathrm{sim}, i}$ denotes the number of simulated galaxies in the $i$th axis ratio bin and $N^{\prime}_{\mathrm{obs},i}$ the weighted number of observed galaxies therein; $\sigma_{i}$ are their respective Poisson uncertainties. Following the modified Neyman's $\chi^2$, when counts of galaxies are zero, a minimum counts of one is enforced to avoid the situation of $\sigma_i=0$. $N_{\mathrm{obs},i,j}$ is the number of observed galaxies in both the $i$th axis ratio bin and the $j$th stellar mass bin. We will clarify the definition of their weights $w_{\mathrm{obs},j}$ when describing the details of the stellar-mass weighting scheme. $N_{{\rm bins},q}=20$ is the number of axis ratio bins with a fixed width of 0.05 and $N_{{\rm bins}, M_\star}=10$ is the number of stellar mass bins with a fixed width of 0.15\,dex. 

It is worth noting that we construct the $\chi^2$ test statistic based on the difference in the number of galaxies between observation and simulation, rather than using the ratio of the number of galaxies in each axis ratio bin to the total number, as done in a previous study by \citetalias{Haslbauer:2022}. This approach is chosen because, in the former case, we can approximate the Poisson counts by a Gaussian distribution, allowing the sum of the test statistics to follow a $\chi^2$ distribution. In contrast, when using ratios, the total number of galaxies itself is a Poisson random variable. This complicates the distribution of the ratios, making it difficult to approximate their sum with a $\chi^2$ distribution.



The different stellar mass distributions of the observed and the simulated galaxies should be taken into account when comparing their axis ratios (see Figure~\ref{fig:mass_distribution}), as galaxy shape is known to depend on stellar mass. We follow the same stellar mass-weighting method in \citetalias{Haslbauer:2022} to correct the axis ratio distribution of the observed galaxies before comparing it with that of the simulated galaxies.  We calculate a weight for each observed galaxy as the number ratio of simulated galaxies to observed galaxies in the corresponding stellar mass bin. Specifically, the weight of an observed galaxy in the $j$th stellar mass bin is defined as $w_{\mathrm{\rm obs},j} = N_{\mathrm{sim},j}/N_{\mathrm{obs},j}$. In the $i$th axis ratio bin, $N^{\prime}_{\mathrm{obs},i}=\sum_{j}{w_{\mathrm{obs},j}N_{\mathrm{obs},i,j}}$ represents the total weight of galaxies therein, a summation of their $w_{{\rm obs},j}$. So, $N^{\prime}_{\mathrm{obs},i}$ is the adjusted number of the observed HSC-SSP galaxies in the axis ratio bin after the stellar-mass weighting, as if the observed galaxies obey the same stellar mass distribution of the TNG50 galaxies. The total weight of all the observed galaxies by definition equals the total number of simulated galaxies.

We then compute the $p$-value for the test statistic $\chi^2$ as $1-$\code{scipy.stats.chi2.cdf($\chi^2$, dof=$N_{{\rm bins},q}-1$)}. In addition, we translate the $p$-value to $n\sigma$ significance as if the test statistic is a Gaussian variable by \code{scipy.stats.norm.ppf($1-p/2$)}, in order to help comparison with previous studies. The statistics for all three bands are summarized in Table~\ref{tab:chisq_tension}. We also include the measurements from \citetalias{Haslbauer:2022} for comparison, even if their definition of $\chi^2$ is different from ours and the upper limit of our stellar mass range is slightly smaller by 0.15\,dex. We find that using their definition of $\chi^2$ results in a minor difference of at most $\sim 10\%$ reduction in our $\chi^2$ values. The small fraction of galaxies in the stellar mass range of $11.5\leq\logM\leq11.65$ has negligible impacts on the statistics. To sum up, the stark difference between their extreme $\chi^2$ ($\gtrsim 200$) and ours cannot be accounted for by the difference in the implementation details. 

\begin{figure*}
     \centering
	\includegraphics[width=1\linewidth]{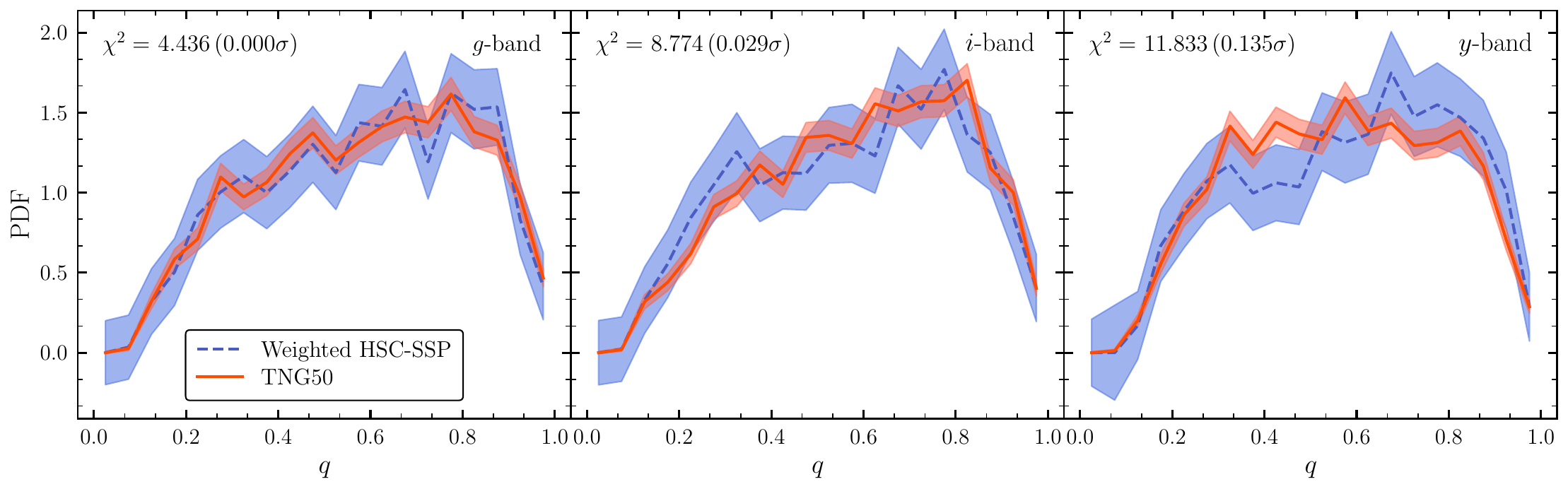}
     \caption{Comparison of the overall axis ratio distributions for the HSC-SSP galaxies and the TNG50 galaxies in the stellar mass range of $10\leq \log (M_{\star}/M_{\sun}) \leq 11.5$. From left to right, we show the comparisons in the $g$, $i$, and $y$ bands, respectively. Solid red lines represent the distribution of the TNG50 galaxies and dashed blue lines represent that of the real HSC-SSP galaxies after stellar mass-weighting. The shaded areas, which share the same colors of their respective lines, enclose the $1\sigma$ Poisson uncertainties calculated following Equations~(\ref{eq:sigma_obs})~and~(\ref{eq:sigma_mod}). The $\chi^2$ statistics and their significance are also shown in the figure and in Table~\ref{tab:chisq_tension}.}
     \label{fig:compare_previous} 
\end{figure*}

Figure~\ref{fig:compare_previous} shows the axis ratio distributions for the TNG50 galaxies (red solid) and the HSC-SSP galaxies after stellar-mass weighting (blue dashed). We find that the stellar-mass weighting has minor impacts on the axis ratio distribution of the HSC-SSP galaxies, because they have similar stellar mass distribution to that of the TNG50 galaxies (see right panel of Figure~\ref{fig:mass_distribution}). The excellent agreement of the axis ratio distributions between the observed and simulated galaxies contradicts the findings of \citetalias{Haslbauer:2022} that TNG50 produces too few thin galaxies.

In addition, we adopt from \citetalias{Haslbauer:2022} the parameter that quantifies the fraction of galaxies with axis ratios smaller than 0.4, $f_{q\leq0.4} = N_{q \leq 0.4}/N_{\mathrm{tot}}$, where $N_{q\leq0.4}$ is the number of galaxies with $q \leq 0.4$ and $N_\mathrm{tot}$ is the total. For the observed HSC-SSP galaxies, the number counts have been adjusted in the aforementioned stellar-mass weighting, and therefore their total number equals that of the TNG50 galaxies. The parameter is not a direct measure of the fraction of intrinsically thin galaxies but is a function of the distribution of intrinsic thickness, in a way that more intrinsically thin galaxies yield larger $f_{q\leq 0.4}$. The uncertainty of $f_{q \leq 0.4}$ for the TNG50 galaxies is calculated as the Gaussian approximation for binomial proportion $\sqrt{f_{q\leq 0.4}(1-f_{q\leq 0.4})/N_{\mathrm{tot}}}$. For the stellar-mass weighted HSC-SSP sample, because the stellar-mass weighting has minor impacts on the $q$ distribution, we can approximate the uncertainty of $f_{q\leq 0.4}$ using the same formula as for the TNG50 galaxies.

\begin{figure*}
     \centering
	\includegraphics[width=0.9\linewidth]{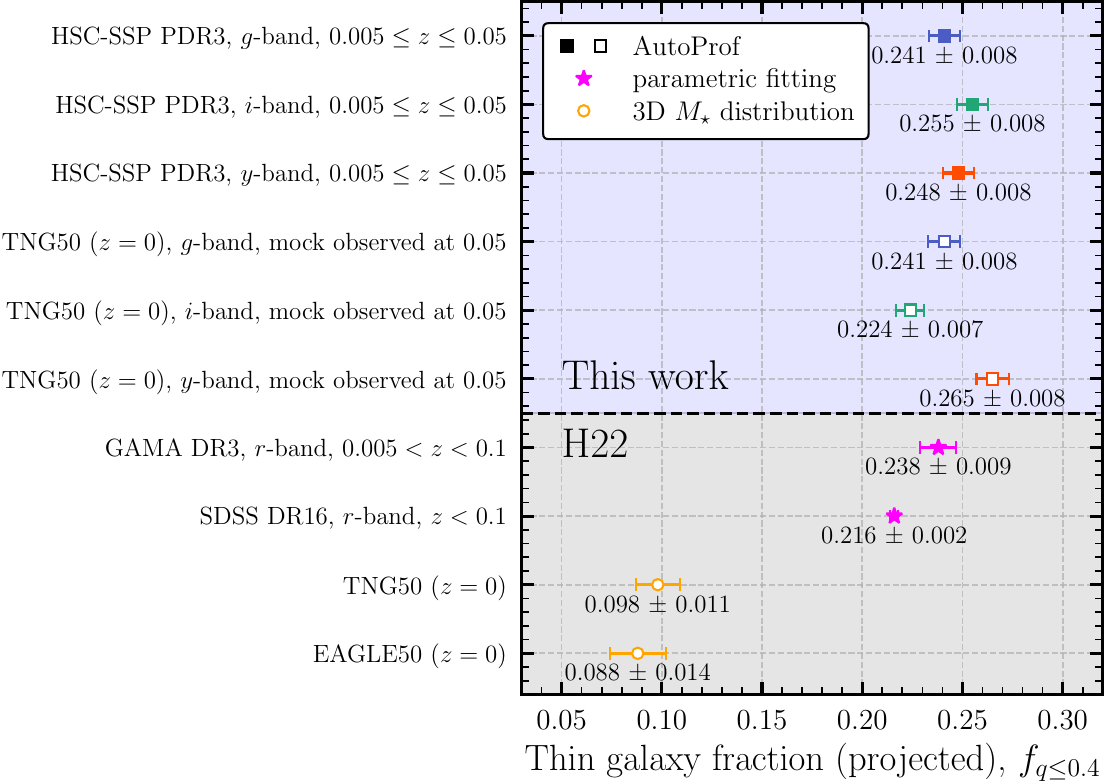}
     \caption{$f_{q\leq 0.4}$ of observed and simulated galaxies in our study and \citetalias{Haslbauer:2022}. The $f_{q\leq 0.4}$ values and their errors from this work and \citetalias{Haslbauer:2022} are shown in the blue and grey shaded areas, respectively. The left text briefly describes the samples, the filters of images, and the redshift ranges. Different shapes of symbol represent the various methods used to measure the axis ratios; open and filled symbols are measurements of simulated galaxies and observed galaxies, respectively. Symbols of the same colors indicate the same filter of (synthetic) images or the same format of simulation data from which the axis ratios were derived. This work uses \code{AutoProf} to measure the axis ratio. The axis ratios of the GAMA sample were derived from single S\'{e}rsic fits using \code{GALFIT}; for the SDSS sample, the axis ratios were obtained via exponential or de~Vaucouleurs fits using the SDSS pipeline. Both were compiled by \citetalias{Haslbauer:2022}. The sky-projected axis ratios of the simulated galaxies in \citetalias{Haslbauer:2022} were derived from their 3D stellar mass distributions. Stellar-mass weighting is applied to observational samples to match the stellar mass distributions of the TNG50 galaxies.}
     \label{fig:frac_smallq} 
\end{figure*}

Figure~\ref{fig:frac_smallq} shows the measurements of $f_{q\leq0.4}$ for various samples of observed and simulated galaxies, after applying the stellar mass weighting to the observational samples. As expected, the $f_{q\leq 0.4}$ of the TNG50 galaxies ($\sim 24\%$) aligns closely with those of the HSC-SSP galaxies ($\sim 25\%$). Moreover, the $f_{q\leq 0.4}$ of the HSC-SSP galaxies ($\sim 25\%$) more or less agrees with the observational results collected by \citetalias{Haslbauer:2022} from other imaging surveys ($\sim 22\%$ for SDSS and $\sim 24\%$ for GAMA). Interestingly, they are all significantly larger than the $f_{q\leq 0.4}$ reported for simulated galaxies ($\sim 10\%$) in \citetalias{Haslbauer:2022}, which implies that the disagreement they found was rooted in their analysis of the simulations. We will discuss more about this in Section~\ref{sec:fair_comp}.

\begin{deluxetable*}{cccCCC}
     \label{tab:chisq_tension}
     \tablecaption{Comparisons of Axis Ratio Distributions between the Observed and the Simulated Galaxies}
     \tablewidth{0pt}
     \tablehead{
     \nocolhead{Works} & 
     \colhead{Sample 1} & 
     \colhead{Sample 2} & 
     \colhead{$\chi^2$} & 
     \colhead{$n\sigma$} &
     \colhead{$p$-value} \\
     \nocolhead{} & \colhead{(1)} & \colhead{(2)} & \colhead{(3)} & \colhead{(4)} & \colhead{(5)}
     }
     \startdata
     This work & HSC-SSP PDR3, $g$ band & TNG50, synthetic $g$ band & $4.44$ & 0.00\sigma & 1.00 \\
     This work & HSC-SSP PDR3, $i$ band & TNG50, synthetic $i$ band & $8.77$ & 0.03\sigma & 0.98 \\
     This work & HSC-SSP PDR3, $y$ band & TNG50, synthetic $y$ band & $11.8$ & 0.13\sigma & 0.89 \\
     \citetalias{Haslbauer:2022} & GAMA DR3, $r$-band & TNG50, 3D stellar mass distribution & $221$ & 12.5\sigma & \nodata \\
     \citetalias{Haslbauer:2022} & SDSS DR16, $r$-band & TNG50, 3D stellar mass distribution & $358$ & 16.9\sigma & \nodata \\
     \enddata
     \tablecomments{Columns~(1)~and~(2): the two samples and the data used for comparison. Column~(3): the $\chi^2$ calculated using Equation~(\ref{eq:chisq_weight}). Column~(4): the $n\sigma$ statistical significance as if the test statistic is a Gaussian variable. Column (5): the $p$-value.}
\end{deluxetable*}

\subsection{Comparison of Axis Ratio Distributions between HSC-SSP and TNG50 at Fixed Stellar Masses}\label{sec:q_mass}

\begin{figure*}
     \centering
	\includegraphics[width=0.9\linewidth]{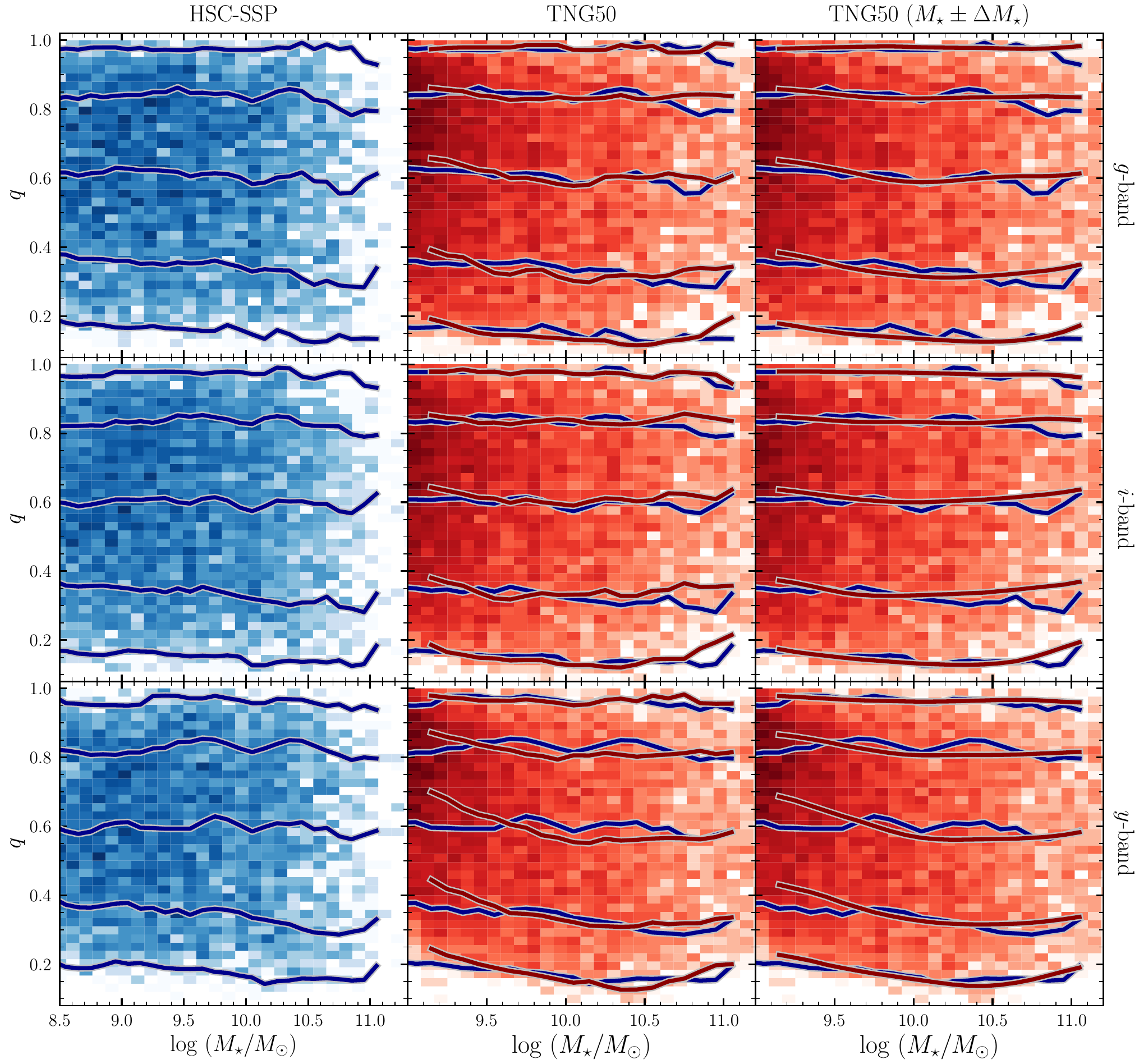}
     \caption{Distributions of galaxies as a function of axis ratio and stellar mass. The left and middle panels show the real and mock HSC-SSP observations (TNG50), respectively. The right panel also displays the mock HSC-SSP observations, but with random perturbations of $0.2$\,dex applied to the stellar masses. Each row, from top to bottom, displays $g$-band, $i$-band, and $y$-band measurements, respectively. Darker colors correspond to a higher density of galaxies in each bin. The solid lines represent the 1st, 16th, 50th, 84th, and 99th percentiles of axis ratios at a given stellar mass. The blue lines in the middle and right panels are replicated percentile lines for the HSC-SSP galaxies in the left panels, to facilitate the comparison with the red ones for the TNG50 galaxies.}
     \label{fig:q_logM} 
\end{figure*}

Building on the remarkable consistency in the overall axis ratio distributions between the observations and TNG50, this section will delve deeper into exploring the dependence of axis ratios on stellar mass over a broader stellar mass range. The $q \textendash \log M_{\star}$ diagram shown in Figure~\ref{fig:q_logM} juxtaposes the galaxies in HSC-SSP and TNG50 with stellar masses of $9.0\leq \logM \leq 11.2$. We truncate the sample at $\logM=11.2$ to ensure enough galaxies at the highest stellar mass for decent sampling of their axis ratio distribution. To simulate the uncertainties in real-world stellar mass estimates, we also randomly perturb the stellar masses of the TNG50 galaxies as suggested by \citet{Rodriguez-Gomez:2019} and show the results in the rightmost column of Figure~\ref{fig:q_logM}. To illustrate the dependence of the axis ratio distributions on stellar mass, we show the 1st, 16th, 50th, 84th, and 99th percentiles of the axis ratios in each stellar mass bin of $0.25$\,dex width. The percentiles of the HSC-SSP and TNG50 galaxies show broad agreement with each other. However, we note that there are noticeable differences at lower stellar masses ($\logM\lesssim 9.5$) and in redder bands ($y$ band). We perform similar comparisons as those in Section~\ref{sec:res_qdis} but in much smaller stellar mass bins and find that the difference between the axis ratio distributions of the HSC-SSP galaxies and the TNG50 galaxies are generally small ($\leq 2\sigma$) except for the $y$-band axis ratios in the stellar mass range $\logM<9.5$, which surges to $>8\sigma$ (Table~\ref{tab:compare_overallq_dins}). Moreover, the statistical significance of the differences at low stellar mass increases monotonically toward longer wavelengths that better probe stellar structures of galaxies, although the increase from $g$ to $i$ band is marginal compared to that from $i$ to $y$ band. The differences in the $y$-band axis ratios are still significant even after smoothing the trend with perturbing the TNG50 stellar masses. 

Overall, the $q_0$\footnote{We use the 1st percentile of the $q$ distribution as a robust approximation of the minimum axis ratio.} shows a subtle (except for in the $y$ band) U-shape dependence on stellar mass for the TNG50 galaxies, which is reminiscent of the one reported by \citet{Sanchez-Janssen:2010}. But, the vertex masses of $\logM \approx 10.4$ is different from $\logM \approx 9.3$ of \citet{Sanchez-Janssen:2010}. We also find that the $q_0$ of the HSC-SSP galaxies has no discernible dependence on stellar masses in all three bands, and therefore there is no local minimum of $q_0$ at $\logM \approx 9$, even with the extended stellar mass range. We suspect that the difference in the approaches to measure axis ratio may be the cause, because \citet{Sanchez-Janssen:2010} used axis ratios measured at a constant surface brightness of 25\,mag\,arcsec$^{-2}$. However, even if we replace our $q$ measurements with those measured at the same surface brightness level as \citet{Sanchez-Janssen:2010}, we are still not able to reproduce the U-shape trend in the HSC-SSP sample. The difference in sample selection may also contribute to the difference in results, as the sample of \citet{Sanchez-Janssen:2010} is closer and spans a smaller distance range (25--80\,Mpc) than ours. The origin of the difference between our results and those of \citet{Sanchez-Janssen:2010} is unclear, and further investigation is beyond the scope of this study.

\begin{deluxetable}{CCCCC}
     \label{tab:compare_overallq_dins}
     \tablecaption{Comparisons of Axis Ratio Distributions as a Function of Stellar Masses}
     \tablewidth{0pt}
     \tablehead{
     \colhead{Filter}& \multicolumn{4}{c}{$\logM$}\\
     \cline{2-5}
     \nocolhead{} & 
     \colhead{9.0--9.5} & 
     \colhead{9.5--10} & 
     \colhead{10.0--10.5} & 
     \colhead{10.5--11.2}
     }
     \startdata
     $g$ & 20.8 (0.9$\sigma$) & 30.0 (1.9$\sigma$) & 30.9 (2.0$\sigma$) & 17.7 (0.6$\sigma$)\\
     $i$ & 28.7 (1.8$\sigma$) & 25.6 (1.5$\sigma$) & 24.2 (1.3$\sigma$) & 22.9 (1.2$\sigma$)\\
     $y$ & 126 (8.4$\sigma$) & 25.4 (1.4$\sigma$) & 26.9 (1.6$\sigma$) & 30.7 (2.0$\sigma$)\\
     \enddata
     \tablecomments{$\chi^2$ and their statistical significance (in parentheses) of the difference between the axis ratios of the HSC-SSP and the TNG50 galaxies in different stellar mass bins observed in the $g$, $i$, and $y$ bands, respectively.}
\end{deluxetable}

\section{Discussion}\label{sec:discussions}
\subsection{The Significance of Equitable Comparisons}\label{sec:fair_comp}

In Section~\ref{sec:res_qdis}, our self-consistent comparisons of HSC-SSP and TNG50 galaxies revealed remarkable agreement between their shapes. Our results contrast strongly with those of \citetalias{Haslbauer:2022}, which reported statistically significant tensions between observed and simulated galaxies---citing a deficit of thin simulated galaxies as a key characteristic of this tension. In this section, we will discuss plausible causes for the discrepancies between our results and theirs.

Most obviously, their comparison used heterogeneous observational measurements of galaxy axis ratios collected from literature and applied a qualitatively different method to measure axis ratios from the simulations. On the contrary, we applied the same method to data of the same format and quality (real and synthetic HSC-SSP images) for the observed sample and the TNG50 galaxies, thanks to the forward modeling of the simulations by \citet{Bottrell:2023}. The approach incorporates various complications that impact the measurements of galaxy shapes, such as stellar population gradients, dust attenuation and scattering, and instrumental effects (PSF smearing and noise), which would otherwise be ignored in measurements based on idealized stellar mass distributions. We note that the observational measurements collected by \citetalias{Haslbauer:2022} agree with both the real HSC-SSP measurements and the mock HSC-SSP measurements of the TNG50 galaxies obtained in this study (Figure~\ref{fig:frac_smallq}), albeit the adopted techniques are different. Only the axis ratios of the simulated galaxies obtained by \citetalias{Haslbauer:2022} consistently stand out as outliers. This leads us to wonder why the method that \citetalias{Haslbauer:2022} applied to the simulations yielded such discrepant results. 

\citetalias{Haslbauer:2022} first extracted the 3D axis lengths of the simulated galaxies defined as $\sqrt{\sum_j m_j r_{j,i}^2}/\sqrt{\sum_j m_j}$ from supplementary catalogs of the corresponding simulations \citep[e.g.,][for TNG]{Genel:2015}, where $m_j$ is the stellar mass of the $j$th particle and $r_{j,i}$ is its $i$th axis coordinate. They then projected an ellipsoid extended by such axis lengths onto a sky plane and built a distribution of projected axis ratios under the assumption of random lines of sight. This approach may not be appropriate for deriving comparable axis ratios to those obtained from the images. Firstly, the assumption of a single ellipsoid for their shape is oversimplified for disk galaxies. It blends all structural components together, therefore the spheroidal components, such as stellar bulges and halos, inevitably contribute to the 3D axis ratio calculations and then to the projected axis ratios at all inclinations. The fact that the 3D axis ratios are defined as the second moment of stellar mass distribution may exaggerate the contamination from the spheroidal components when compared with galaxy shapes measured from optical images, because (1) stellar bulges and halos are often fainter than disks at fixed stellar mass density due to their higher optical mass-to-light ratios and (2) they contribute substantially to the intrinsic flatness $q_0$---especially the stellar halo particles far away from the disk plane. In the EAGLE simulation, \citet{deGraaff:2022} found that galaxy axis ratio distributions measured from projected stellar mass maps are consistent with those from their $r$-band images, which suggests that the former may not be the primary cause and the latter plays a more dominant role. 

Our method of fitting an isophote in the outskirt of the projected galaxy images strives to avoid the center. Also, it does not reach the surface brightness level of stellar halos \citep{Merritt:2016}. This approach provides reliable measurements of the projected shape of disks, except for edge-on galaxies where bulges inevitably come into play. It makes sense that the axis ratios obtained via parametric fitting (i.e., fitting 2D parametric surface brightness models to images) are broadly consistent with ours in terms of $f_{q\leq 0.4}$ (Figure~\ref{fig:frac_smallq}), because both our method and parametric fitting are essentially isophote-based methods. Still, differences in the details remain. Parametric fitting assumes model isophotes to be concentric and to have the same shape, so the best-fit axis ratio is the average axis ratio of the overall flux distribution and would be biased by central components that have a different shape such as bulges. Our method, in contrast, is more robust against such bias.

In summary, the significant disagreement in galaxy axis ratios between the cosmological simulations and the observations reported by \citetalias{Haslbauer:2022} can be attributed to the fundamentally different methods applied to obtain the measurements from both sides. Crafting synthetic images that are compatible with observational data enables the application of consistent methods to extract the same observables for comparison with observations. Our study proves such an exercise pivotal by presenting robust and distinct results from those obtained via a different approach.

\subsection{The Low-mass Thicker Galaxies in TNG50}\label{sec:low-mass-shape}

In Section~\ref{sec:q_mass}, we noted that the TNG50 sample has axis ratio distributions skewed toward larger $q$ compared with those of the HSC-SSP galaxies in the low-mass regime, while the differences are much less significant at higher stellar masses. This implies that the low-mass galaxies in TNG50 are intrinsically rounder/thicker than their observed counterparts. Spurious heating, the gravitational scattering between stellar and dark matter particles due to finite resolution, is known to increase stellar velocity dispersions and the 3D thicknesses of galaxies in cosmological simulations \citep{Ludlow:2021, Wilkinson:2023}. Spurious heating should have the most profound effects on low-mass galaxies comprising smaller numbers of stellar and dark matter particles \citep{Ludlow+2023}. This can explain the rounder shapes of the low-mass TNG50 galaxies than the observations.

It is worth noting that the generation of the TNG50 synthetic images involves post-processing of the dust content in the simulated galaxies. Because TNG does not explicitly track dust, an empirical model motivated by dust-to-gas ratios in nearby galaxies \citep{Remy-Ruyer+2014} was used to infer the dust mass from the gas mass in each gas cell based on its metallicity and temperature \citep{Popping+2022,Bottrell:2023}. \citet{Rodriguez-Gomez:2019} showed that whether modeling dust radiative transfer or not modeling dust at all has a negligible effect on observed median ellipticity derived from synthetic images of simulated galaxies (their Figure~7). Therefore, we argue that the choice of dust post-processing model cannot have such a significant effect to account for the difference between the observed and simulated median $q$ of the low-mass galaxies (Figure~\ref{fig:q_logM}).

\begin{figure}
     \centering
	\includegraphics[width=1\linewidth]{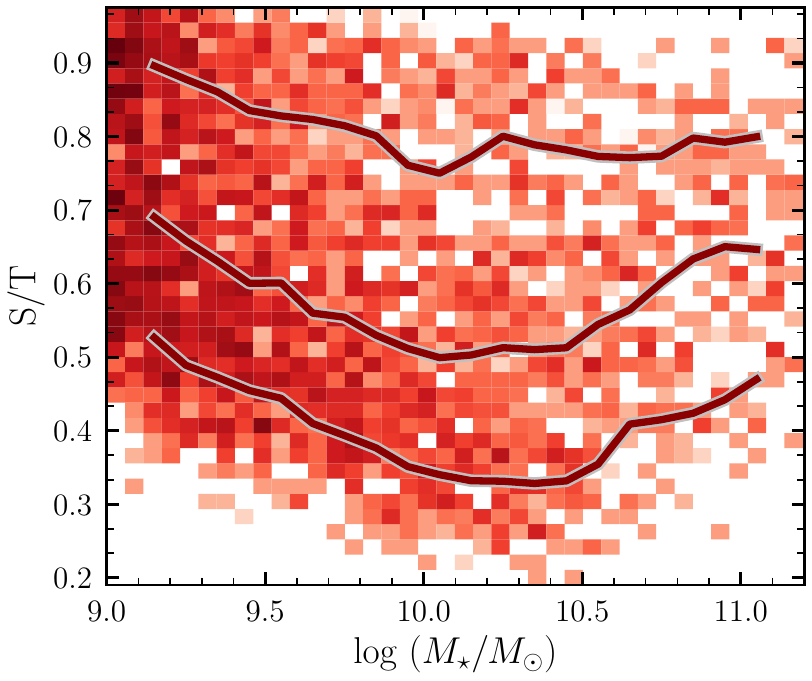}
     \caption{The relation between S/T and stellar mass in TNG50. The three red lines mark the 16th, 50th, and 84th percentiles of S/T.
     }
     \label{fig:st_logM} 
\end{figure}

In addition, as highlighted by D. Xu et al. (2024, in preparation), axis ratio measurements for edge-on galaxies are affected by kinematically hot structures such as stellar bulges and other spheroidal components. Consequently, variations in structural compositions of galaxies may play an important role in shaping the dependence of axis ratios on stellar masses. The U-shape dependence of $q_0$ on stellar mass is reminiscent of that of S/T on stellar mass in TNG, first reported by \citet{Tacchella:2019} in TNG100 and reproduced in Figure~\ref{fig:st_logM} for TNG50. A similar trend was also found in the EAGLE simulation \citep{Clauwens:2018}. The median S/T reaches its minimum at $10< \logM < 10.5$ in both TNG and EAGLE. The coincidence of the vertex stellar masses of the $q_0$--$\log M_{\star}$ and S/T--$\log M_{\star}$ relations and the stellar mass where early-type galaxies start to dominate in TNG \citep{Huertas-Company+2019} leads us to investigate whether the stellar mass dependence of galaxy structural compositions also contributes to the curvature in the $q_0$--$\log M_{\star}$ relation of TNG50. 

\begin{figure*}
     \centering
	\includegraphics[width=1\linewidth]{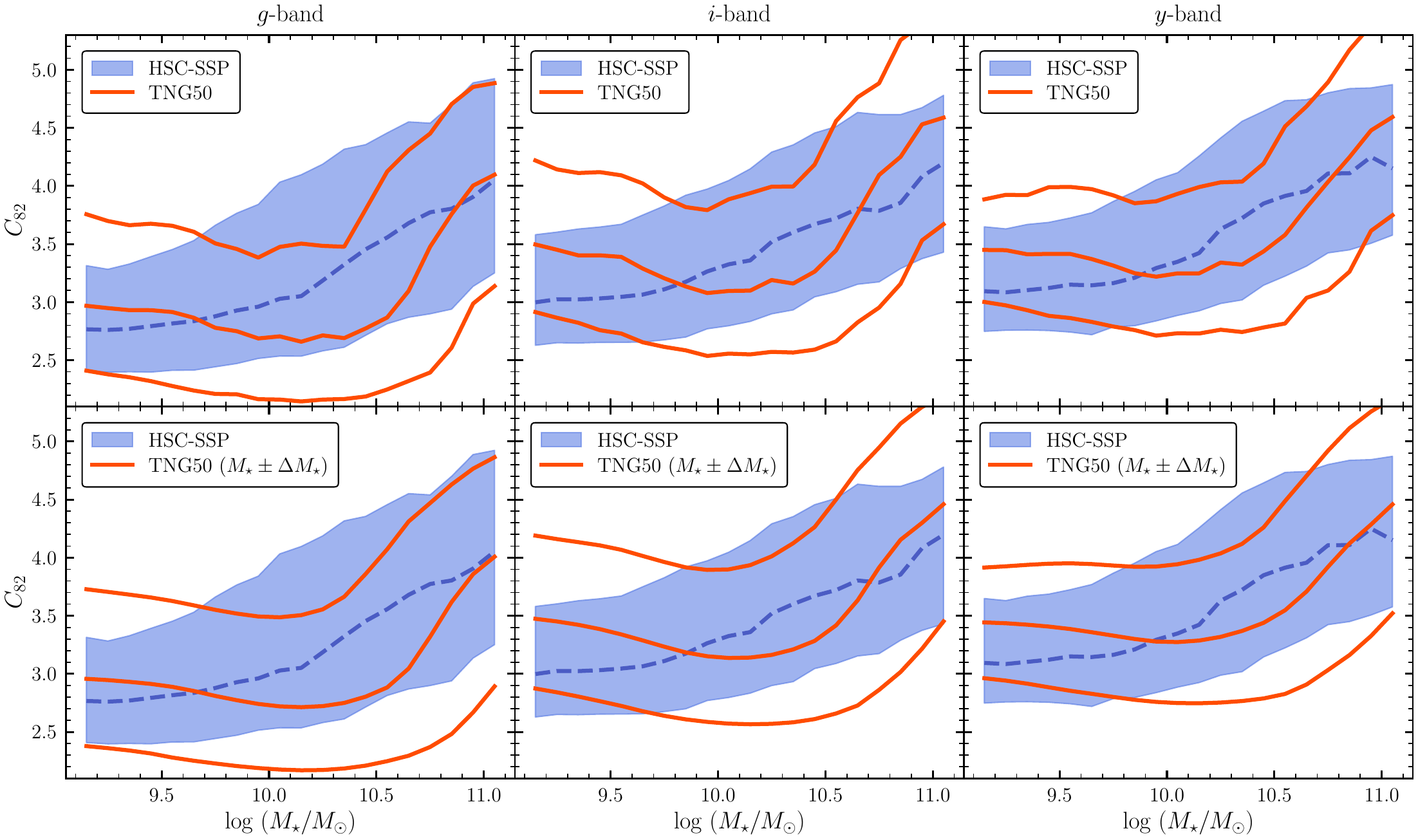}
     \caption{The relations between concentration index and stellar mass of the HSC-SSP galaxies and the TNG50 galaxies. In the top row, solid lines represent the 16th, 50th, and 84th percentiles of $C_{82}$ for the TNG50 galaxies, and the shaded regions and dashed lines mark the same percentiles for the HSC-SSP galaxies. The bottom row follows the same convention, except that the stellar masses of the TNG50 galaxies are randomly perturbed as in Figure~\ref{fig:q_logM}.            
     \label{fig:C82_mass}}
\end{figure*}

In Figure~\ref{fig:C82_mass}, we compare the $C_{82}$--$\log M_{\star}$ relation of the observed galaxies and the TNG50 galaxies. The $C_{82}$ is commonly recognized as an economical substitute for the \citet{Sersic:1968} index and bulge-to-total ratio \citep[e.g.,][]{Gadotti:2009}. We find broad agreement between the two as well as some interesting differences. The median relations of the TNG50 galaxies exhibit a minimum at a similar stellar mass as the aforementioned U-shape relations, while the median relations of the HSC-SSP galaxies are monotonic. Because it is unclear how the spheroidal components quantitatively contribute to the $q_0$ measurements, the absolute difference of $C_{82}$ at fixed stellar masses between the observations and TNG50 is irrelevant for interpreting the shape of the $q_0$--$\log M_{\star}$ relations. What matters is the relative difference of the trends with stellar mass between them. 

Toward lower stellar masses, the TNG50 relations show a less significant decrease in $C_{82}$ compared to the observational trends in all three bands. Toward higher stellar masses, the TNG50 galaxies exhibit a more prominent increase in $C_{82}$ than the observed ones in all three bands. These differences persist even after smoothing the trends by randomly perturbing the stellar masses of the simulated galaxies. This is also true for the comparison of concentrations and bulge statistics between TNG100 galaxies and real galaxies observed in the Panoramic Survey Telescope and Rapid Response System (Pan-STARRS\@; \citep{Chambers+2016} $i$-band images at $z\approx 0.05$ \citep{Rodriguez-Gomez:2019}. Assuming that more prominent spheroidal components at a given stellar mass lead to larger $q_0$ estimates of galaxies in the same sample, the difference in the $q_0$--$\log M_{\star}$ relations is compatible with the more curved TNG50 $q_0$--$\log M_{\star}$ relations than the HSC-SSP ones. It also explains why the U-shape is most noticeable in the reddest $y$ band, where the spheroidal components should be most prominent in flux (as seen in the elevation of $C_{82}$ from the $g$ to $y$ band; Figure~\ref{fig:C82_mass}), because they are mostly quenched structures in massive galaxies. However, we argue that this cannot be the dominant factor for the low-mass thicker galaxies in TNG50, otherwise we should observe a similar degree of mismatch between the observations and TNG50 at the high-mass end and should expect to observe thicker low-mass TNG50 galaxies in the $g$ and $i$ bands as well. 

In conclusion, we suggest that the numerical effects of spurious heating are the main cause for the low-mass thicker galaxies in TNG50. The larger concentration of the TNG50 galaxies may contribute to the difference in axis ratios, but we argue that this should only have a minor effect.


\subsection{Difference in Galaxy Concentration between Observations and TNG50}

The difference in concentration index between the observations and TNG50 is intriguing. As mentioned in Section~\ref{sec:low-mass-shape}, our results are consistent with the trends reported by \citet{Rodriguez-Gomez:2019}, despite their different definition of total fluxes from ours. However, \citet{Guzman-Ortega:2023} reported different results that TNG50 galaxies have systematically larger average concentration than galaxies observed in the Kilo-Degree Survey (KiDS\@; \citealp{deJong:2013}) in the range of $8.5 \leq \logM \leq 11$ in a similar redshift range of $z<0.05$. Both studies performed similar research using synthetic images and reported broad agreement between the observations and the simulations as our work, and yet high-order differences persist. Difference in image qualities such as survey depth and resolution may contribute to the high-order difference in the trends.

The difference between observations and simulations at the low-mass end can be attributed to poorer numerical resolution. In addition, \citet{Rodriguez-Gomez:2019} showed that different dust models yielded $\sim 0.1$\,dex variations in the mock observed $i$-band concentration index of simulated galaxies at $\logM \approx 9.5$ (their Figure~6). Understanding the difference at the high-mass end requires more dedicated effort. Dust modeling in generating the synthetic images has a negligible effect since massive galaxies are mostly gas-poor \citep{Rodriguez-Gomez:2019}. \citet{Guzman-Ortega:2023} found slightly higher merger rates in TNG50 than in KiDS galaxies using both random forest and asymmetry-based classifications, which seems a plausible explanation for the higher concentrations in TNG50 galaxies. However, they also showed that merger and non-merger TNG50 galaxies have similar $r$-band concentrations. Moreover, classification of mergers suffers from incompleteness and impurity issues \citep[e.g.,][]{Bickley+2021, Bottrell+2022}. The uncertainties in stellar mass estimates of the observed galaxies could also contribute to the difference in the $C_{82}$--$\log M_{\star}$ relations \citep{Rodriguez-Gomez:2019}. We strive to emulate such uncertainties by randomly perturbing the stellar masses of the simulated galaxies and find that the difference between TNG50 and the observations remains significant despite the smoother trends in TNG50 (Figure~\ref{fig:C82_mass}). However, we caution that this approach does not account for the inclination dependent uncertainties in stellar mass estimates.

\section{Summary}\label{sec:summary}

We perform rigorous comparisons of galaxy shapes between the TNG50 simulation and observational data. To guarantee the comparability of the measurements from both sides, we make use of the synthetic images of the TNG50 snapshot 99 ($z=0$) mock observed at $z=0.05$ and real galaxy images from HSC-SSP PDR3 \texttt{WIDE} within a similar redshift range. We apply the same techniques to both datasets for generating object masks using \code{sep} and measuring axis ratios using \code{AutoProf}. Our findings are summarized as follows.
\begin{itemize}
    \item We find excellent agreement ($\lesssim 0.1\sigma$) in the overall galaxy axis ratio distribution between the TNG50 simulation and the HSC-SSP observations in the stellar mass range $10.0 \leq \logM \leq 11.5$. Our results contrast with the significant tensions ($\gtrsim 10\sigma$) reported by \citetalias{Haslbauer:2022}. The stark difference between our results and theirs can be ascribed to our more self-consistent methodology that delivers a more equitable comparison.
    \item We find that low-mass galaxies ($\logM \lesssim 9.5$) in TNG50 are thicker than their observed counterparts. We suggest that spurious dynamical heating of simulated stellar components by dark matter particles is the main cause, instead of unrealistic physics. 
    \item We find that the TNG50 galaxies are more concentrated than their observed counterparts at the low- and high-mass end of the regime we probe ($\logM = 9 \text{--}11.2$) and are less concentrated at intermediate stellar masses. This may contribute to but should not be consequential to the difference in the $q_0$--$\log M_{\star}$ relations between the observation and the TNG50 simulation, especially at the low-mass end. 
\end{itemize}

Our study highlights the importance of conducting mock observations of simulations and employing consistent methods to extract measurements for comparison with observational results. The minor ($\lesssim 0.1\sigma$) difference in the overall axis ratio distributions between observations and the TNG50 simulation adds to the success of TNG in terms of reproducing realistic galaxy morphologies \citep[e.g.,][]{Rodriguez-Gomez:2019, Guzman-Ortega:2023, Eisert2024contrastive}. Our results call into question the previous claims that $\Lambda$CDM produced too few thin galaxies compared to observations (e.g., \citealp{Kormendy:2010, vandeSande:2019}; \citetalias{Haslbauer:2022}). Nevertheless, we also reveal differences in both the $q_0$--$\log M_{\star}$ and the $C_{82}$--$\log M_{\star}$ relations between the TNG50 simulation and the observation. This emphasizes the necessity to perform detailed comparisons of galaxy morphologies as a function of galaxy properties in addition to the overall distributions \citep{Huertas-Company+2019, Eisert2024contrastive}. We discuss the possible origins of the difference in galaxy concentrations between the simulation and the observations. It will require further investigation to ascertain the primary cause.

\begin{acknowledgments}
H.G. thanks the support from the Pan-STARRS project, which is supported by NASA 80NSSC21K1572 and by the State of Hawaii. C.B. gratefully acknowledges the support of the Forrest Research Foundation. H.M.Y. was partially supported by JSPS KAKENHI Grant Number JP22K14072 and the Research Fund for International Young Scientists of NSFC (11950410492).

We are deeply grateful for the opportunity to observe the Universe from Mauna Kea, which has always held a very significant cultural role within the indigenous Hawaiian community.

The TNG50 simulation was run with compute time granted by the Gauss Centre for Supercomputing (GCS) under Large-Scale Projects GCS-DWAR on the GCS share of the supercomputer Hazel Hen at the High Performance Computing Center Stuttgart (HLRS).

The Hyper Suprime-Cam (HSC) collaboration includes the astronomical communities of Japan and Taiwan, and Princeton University. The HSC instrumentation and software were developed by the National Astronomical Observatory of Japan (NAOJ), the Kavli Institute for the Physics and Mathematics of the Universe (Kavli IPMU), the University of Tokyo, the High Energy Accelerator Research Organization (KEK), the Academia Sinica Institute for Astronomy and Astrophysics in Taiwan (ASIAA), and Princeton University. Funding was contributed by the FIRST program from the Japanese Cabinet Office, the Ministry of Education, Culture, Sports, Science and Technology (MEXT), the Japan Society for the Promotion of Science (JSPS), Japan Science and Technology Agency (JST), the Toray Science Foundation, NAOJ, Kavli IPMU, KEK, ASIAA, and Princeton University.

Funding for SDSS-III has been provided by the Alfred P. Sloan Foundation, the Participating Institutions, the National Science Foundation, and the U.S. Department of Energy Office of Science. The SDSS-III website is \url{http://www.sdss3.org/}. SDSS-III is managed by the Astrophysical Research Consortium for the Participating Institutions of the SDSS-III Collaboration including the University of Arizona, the Brazilian Participation Group, Brookhaven National Laboratory, University of Cambridge, Carnegie Mellon University, University of Florida, the French Participation Group, the German Participation Group, Harvard University, the Instituto de Astrofisica de Canarias, the Michigan State/Notre Dame/JINA Participation Group, Johns Hopkins University, Lawrence Berkeley National Laboratory, Max Planck Institute for Astrophysics, Max Planck Institute for Extraterrestrial Physics, New Mexico State University, New York University, Ohio State University, Pennsylvania State University, University of Portsmouth, Princeton University, the Spanish Participation Group, University of Tokyo, University of Utah, Vanderbilt University, University of Virginia, University of Washington, and Yale University.  

GAMA is a joint European-Australasian project based around a spectroscopic campaign using the Anglo-Australian Telescope. The GAMA input catalog is based on data taken from the Sloan Digital Sky Survey and the UKIRT Infrared Deep Sky Survey. Complementary imaging of the GAMA regions is being obtained by a number of independent survey programs, including GALEX MIS, VST KiDS, VISTA VIKING, WISE, Herschel-ATLAS, GMRT, and ASKAP, providing UV to radio coverage. GAMA is funded by the STFC (UK), the ARC (Australia), the AAO, and the participating institutions. The GAMA website is \url{http://www. gama-survey.org/}.
\end{acknowledgments}


\appendix

\section{The Effect of the Resolution Cut on Completeness of the Axis Ratio Distribution}\label{appendix:rescut}

\begin{figure*}
     \centering
	\includegraphics[width=1\linewidth]{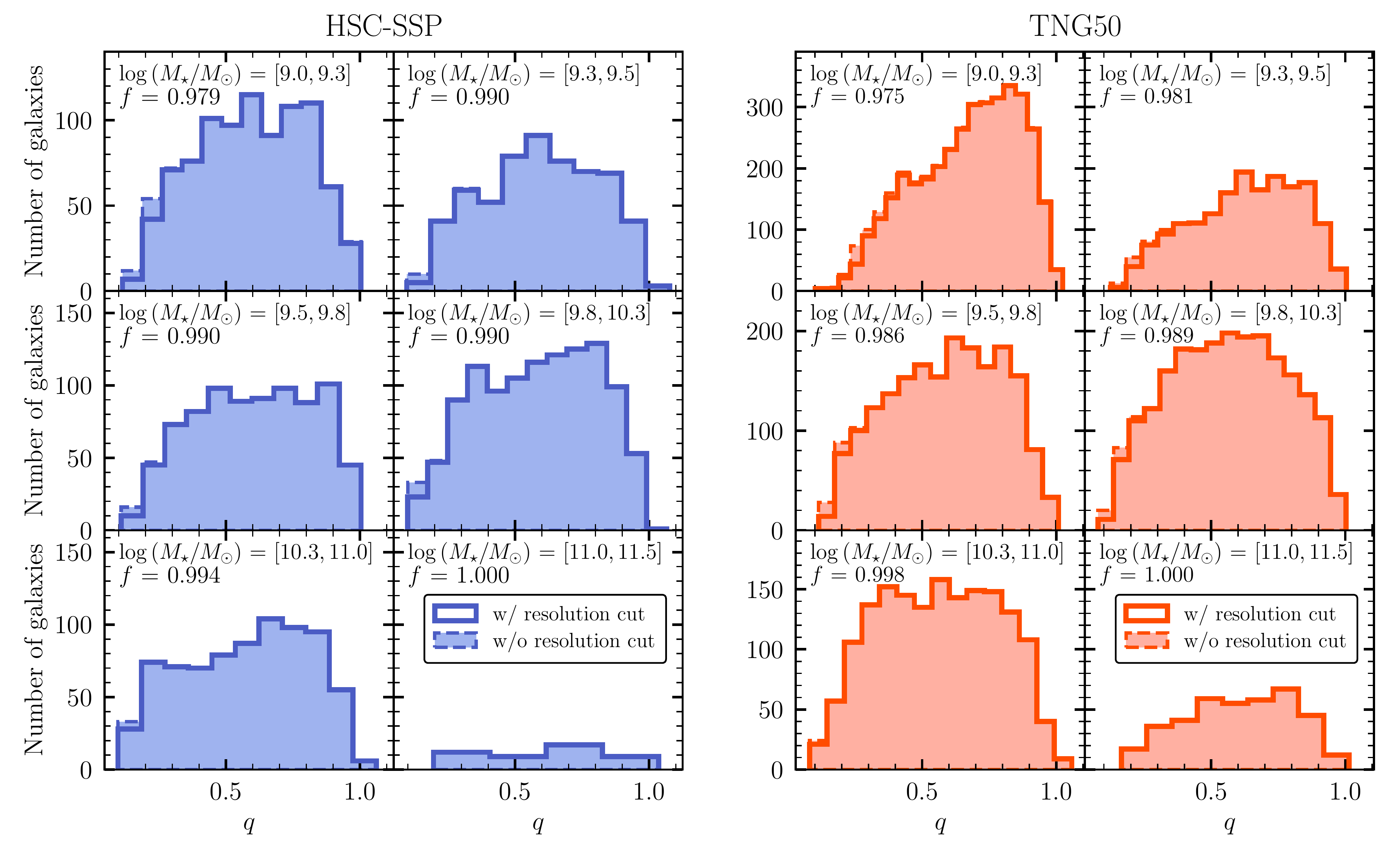}
     \caption{The distributions of $y$-band axis ratios of the HSC-SSP sample (left) and the TNG50 sample (right) with resolution cut (open solid) and without resolution cut (filled dashed) in different stellar mass bins. Stellar mass increases from left to right and from top to bottom in the range $9 \leq \log (M_\star/M_\sun) \leq 11.5$. The parameter $f$ is the ratio of the number of galaxies after the resolution cut to before the resolution cut in each stellar mass bin.}
     \label{fig:dis_q_rescut_bins_obs} 
\end{figure*}

To minimize the PSF smearing effects on the measurements of axis ratios, we excluded the measurements with isophotal semi-minor axis length smaller than $1.5\times$ the PSF FWHM. We use the $y$-band sample to demonstrate the effect of the resolution cut on the completeness of the axis ratio distributions because the $y$-band images in HSC-SSP have the worst seeing. Figure~\ref{fig:dis_q_rescut_bins_obs} illustrates the axis ratio distributions in various stellar mass bins for both the HSC-SSP galaxies and the TNG50 galaxies, with and without the resolution cut. Our findings indicate that the resolution cut has a negligible impact on the completeness of the axis ratio distribution in both samples. Consistent with expectations, the resolution cut primarily affects the least massive (smallest in size) and nearly edge-on galaxies (shortest minor axis). The galaxies are close enough ($z\leq 0.05$) to be sufficiently resolved.

\bibliography{tng50_hsc}{}

\begin{thebibliography}{}
\expandafter\ifx\csname natexlab\endcsname\relax\def\natexlab#1{#1}\fi
\providecommand{\url}[1]{\href{#1}{#1}}
\providecommand{\dodoi}[1]{doi:~\href{http://doi.org/#1}{\nolinkurl{#1}}}
\providecommand{\doeprint}[1]{\href{http://ascl.net/#1}{\nolinkurl{http://ascl.net/#1}}}
\providecommand{\doarXiv}[1]{\href{https://arxiv.org/abs/#1}{\nolinkurl{https://arxiv.org/abs/#1}}}

\bibitem[{{Ahn} {et~al.}(2014){Ahn}, {Alexandroff}, {Allende Prieto}, {Anders}, {Anderson}, {Anderton}, {Andrews}, {Aubourg}, {Bailey}, {Bastien}, \& et~al.}]{Ahn:2014}
{Ahn}, C.~P., {Alexandroff}, R., {Allende Prieto}, C., {et~al.} 2014, \apjs, 211, 17, \dodoi{10.1088/0067-0049/211/2/17}

\bibitem[{{Ahumada} {et~al.}(2020){Ahumada}, {Allende Prieto}, {Almeida}, {Anders}, {Anderson}, {Andrews}, {Anguiano}, {Arcodia}, {Armengaud}, {Aubert}, \& et~al.}]{Ahumada:2020}
{Ahumada}, R., {Allende Prieto}, C., {Almeida}, A., {et~al.} 2020, \apjs, 249, 3, \dodoi{10.3847/1538-4365/ab929e}

\bibitem[{{Aihara} {et~al.}(2018){Aihara}, {Armstrong}, {Bickerton}, {Bosch}, {Coupon}, {Furusawa}, {Hayashi}, {Ikeda}, {Kamata}, {Karoji}, {Kawanomoto}, {Koike}, {Komiyama}, {Lang}, {Lupton}, {Mineo}, {Miyatake}, {Miyazaki}, {Morokuma}, {Obuchi}, {Oishi}, {Okura}, {Price}, {Takata}, {Tanaka}, {Tanaka}, {Tanaka}, {Uchida}, {Uraguchi}, {Utsumi}, {Wang}, {Yamada}, {Yamanoi}, {Yasuda}, {Arimoto}, {Chiba}, {Finet}, {Fujimori}, {Fujimoto}, {Furusawa}, {Goto}, {Goulding}, {Gunn}, {Harikane}, {Hattori}, {Hayashi}, {He{\l}miniak}, {Higuchi}, {Hikage}, {Ho}, {Hsieh}, {Huang}, {Huang}, {Imanishi}, {Iwata}, {Jaelani}, {Jian}, {Kashikawa}, {Katayama}, {Kojima}, {Konno}, {Koshida}, {Kusakabe}, {Leauthaud}, {Lee}, {Lin}, {Lin}, {Mandelbaum}, {Matsuoka}, {Medezinski}, {Miyama}, {Momose}, {More}, {More}, {Mukae}, {Murata}, {Murayama}, {Nagao}, {Nakata}, {Niida}, {Niikura}, {Nishizawa}, {Oguri}, {Okabe}, {Ono}, {Onodera}, {Onoue}, {Ouchi}, {Pyo}, {Shibuya}, {Shimasaku}, {Simet}, {Speagle}, {Spergel}, {Strauss}, {Sugahara}, {Sugiyama}, {Suto}, {Suzuki}, {Tait}, {Takada}, {Terai}, {Toba}, {Turner}, {Uchiyama}, {Umetsu}, {Urata}, {Usuda}, {Yeh}, \& {Yuma}}]{Aihara:2018b}
{Aihara}, H., {Armstrong}, R., {Bickerton}, S., {et~al.} 2018, \pasj, 70, S8, \dodoi{10.1093/pasj/psx081}

\bibitem[{{Aihara} {et~al.}(2019){Aihara}, {AlSayyad}, {Ando}, {Armstrong}, {Bosch}, {Egami}, {Furusawa}, {Furusawa}, {Goulding}, {Harikane}, {Hikage}, {Ho}, {Hsieh}, {Huang}, {Ikeda}, {Imanishi}, {Ito}, {Iwata}, {Jaelani}, {Kakuma}, {Kawana}, {Kikuta}, {Kobayashi}, {Koike}, {Komiyama}, {Li}, {Liang}, {Lin}, {Luo}, {Lupton}, {Lust}, {MacArthur}, {Matsuoka}, {Mineo}, {Miyatake}, {Miyazaki}, {More}, {Murata}, {Namiki}, {Nishizawa}, {Oguri}, {Okabe}, {Okamoto}, {Okura}, {Ono}, {Onodera}, {Onoue}, {Osato}, {Ouchi}, {Shibuya}, {Strauss}, {Sugiyama}, {Suto}, {Takada}, {Takagi}, {Takata}, {Takita}, {Tanaka}, {Terai}, {Toba}, {Uchiyama}, {Utsumi}, {Wang}, {Wang}, \& {Yamada}}]{Aihara:2019}
{Aihara}, H., {AlSayyad}, Y., {Ando}, M., {et~al.} 2019, \pasj, 71, 114, \dodoi{10.1093/pasj/psz103}

\bibitem[{{Aihara} {et~al.}(2022){Aihara}, {AlSayyad}, {Ando}, {Armstrong}, {Bosch}, {Egami}, {Furusawa}, {Furusawa}, {Harasawa}, {Harikane}, {Hsieh}, {Ikeda}, {Ito}, {Iwata}, {Kodama}, {Koike}, {Kokubo}, {Komiyama}, {Li}, {Liang}, {Lin}, {Lupton}, {Lust}, {MacArthur}, {Mawatari}, {Mineo}, {Miyatake}, {Miyazaki}, {More}, {Morishima}, {Murayama}, {Nakajima}, {Nakata}, {Nishizawa}, {Oguri}, {Okabe}, {Okura}, {Ono}, {Osato}, {Ouchi}, {Pan}, {Plazas Malag{\'o}n}, {Price}, {Reed}, {Rykoff}, {Shibuya}, {Simunovic}, {Strauss}, {Sugimori}, {Suto}, {Suzuki}, {Takada}, {Takagi}, {Takata}, {Takita}, {Tanaka}, {Tang}, {Taranu}, {Terai}, {Toba}, {Turner}, {Uchiyama}, {Vijarnwannaluk}, {Waters}, {Yamada}, {Yamamoto}, \& {Yamashita}}]{Aihara:2022}
---. 2022, \pasj, 74, 247, \dodoi{10.1093/pasj/psab122}

\bibitem[{{Bah{\'e}} {et~al.}(2017){Bah{\'e}}, {Barnes}, {Dalla Vecchia}, {Kay}, {White}, {McCarthy}, {Schaye}, {Bower}, {Crain}, {Theuns}, {Jenkins}, {McGee}, {Schaller}, {Thomas}, \& {Trayford}}]{Bahe:2017}
{Bah{\'e}}, Y.~M., {Barnes}, D.~J., {Dalla Vecchia}, C., {et~al.} 2017, \mnras, 470, 4186, \dodoi{10.1093/mnras/stx1403}

\bibitem[{{Barbary} {et~al.}(2016){Barbary}, {Boone}, {McCully}, {Craig}, {Deil}, \& {Rose}}]{Barbary:2016}
{Barbary}, K., {Boone}, K., {McCully}, C., {et~al.} 2016, {Kbarbary/Sep: V1.0.0}, v1.0.0, Zenodo,  Zenodo, \dodoi{10.5281/zenodo.159035}

\bibitem[{{Bickley} {et~al.}(2021){Bickley}, {Bottrell}, {Hani}, {Ellison}, {Teimoorinia}, {Yi}, {Wilkinson}, {Gwyn}, \& {Hudson}}]{Bickley+2021}
{Bickley}, R.~W., {Bottrell}, C., {Hani}, M.~H., {et~al.} 2021, \mnras, 504, 372, \dodoi{10.1093/mnras/stab806}

\bibitem[{{Binggeli}(1980)}]{Binggeli:1980}
{Binggeli}, B. 1980, \aap, 82, 289

\bibitem[{{Bottrell} {et~al.}(2022){Bottrell}, {Hani}, {Teimoorinia}, {Patton}, \& {Ellison}}]{Bottrell+2022}
{Bottrell}, C., {Hani}, M.~H., {Teimoorinia}, H., {Patton}, D.~R., \& {Ellison}, S.~L. 2022, \mnras, 511, 100, \dodoi{10.1093/mnras/stab3717}

\bibitem[{{Bottrell} {et~al.}(2017{\natexlab{a}}){Bottrell}, {Torrey}, {Simard}, \& {Ellison}}]{Bottrell:2017a}
{Bottrell}, C., {Torrey}, P., {Simard}, L., \& {Ellison}, S.~L. 2017{\natexlab{a}}, \mnras, 467, 1033, \dodoi{10.1093/mnras/stx017}

\bibitem[{{Bottrell} {et~al.}(2017{\natexlab{b}}){Bottrell}, {Torrey}, {Simard}, \& {Ellison}}]{Bottrell:2017b}
---. 2017{\natexlab{b}}, \mnras, 467, 2879, \dodoi{10.1093/mnras/stx276}

\bibitem[{{Bottrell} {et~al.}(2019){Bottrell}, {Hani}, {Teimoorinia}, {Ellison}, {Moreno}, {Torrey}, {Hayward}, {Thorp}, {Simard}, \& {Hernquist}}]{Bottrell2019realsim}
{Bottrell}, C., {Hani}, M.~H., {Teimoorinia}, H., {et~al.} 2019, \mnras, 490, 5390, \dodoi{10.1093/mnras/stz2934}

\bibitem[{{Bottrell} {et~al.}(2024){Bottrell}, {Yesuf}, {Popping}, {Omori}, {Tang}, {Ding}, {Pillepich}, {Nelson}, {Eisert}, {Gao}, {Goulding}, {Kalita}, {Luo}, {Greene}, {Shi}, \& {Silverman}}]{Bottrell:2023}
{Bottrell}, C., {Yesuf}, H.~M., {Popping}, G., {et~al.} 2024, \mnras, 527, 6506, \dodoi{10.1093/mnras/stad2971}

\bibitem[{{Bruzual} \& {Charlot}(2003)}]{BruzualCharlot2003}
{Bruzual}, G., \& {Charlot}, S. 2003, \mnras, 344, 1000, \dodoi{10.1046/j.1365-8711.2003.06897.x}

\bibitem[{{Camps} \& {Baes}(2020)}]{Camps&Baes:2020}
{Camps}, P., \& {Baes}, M. 2020, Astronomy and Computing, 31, 100381, \dodoi{10.1016/j.ascom.2020.100381}

\bibitem[{{Chabrier}(2003)}]{Chabrier:2003}
{Chabrier}, G. 2003, \pasp, 115, 763, \dodoi{10.1086/376392}

\bibitem[{{Chambers} {et~al.}(2016){Chambers}, {Magnier}, {Metcalfe}, {Flewelling}, {Huber}, {Waters}, {Denneau}, {Draper}, {Farrow}, {Finkbeiner}, {Holmberg}, {Koppenhoefer}, {Price}, {Rest}, {Saglia}, {Schlafly}, {Smartt}, {Sweeney}, {Wainscoat}, {Burgett}, {Chastel}, {Grav}, {Heasley}, {Hodapp}, {Jedicke}, {Kaiser}, {Kudritzki}, {Luppino}, {Lupton}, {Monet}, {Morgan}, {Onaka}, {Shiao}, {Stubbs}, {Tonry}, {White}, {Ba{\~n}ados}, {Bell}, {Bender}, {Bernard}, {Boegner}, {Boffi}, {Botticella}, {Calamida}, {Casertano}, {Chen}, {Chen}, {Cole}, {Deacon}, {Frenk}, {Fitzsimmons}, {Gezari}, {Gibbs}, {Goessl}, {Goggia}, {Gourgue}, {Goldman}, {Grant}, {Grebel}, {Hambly}, {Hasinger}, {Heavens}, {Heckman}, {Henderson}, {Henning}, {Holman}, {Hopp}, {Ip}, {Isani}, {Jackson}, {Keyes}, {Koekemoer}, {Kotak}, {Le}, {Liska}, {Long}, {Lucey}, {Liu}, {Martin}, {Masci}, {McLean}, {Mindel}, {Misra}, {Morganson}, {Murphy}, {Obaika}, {Narayan}, {Nieto-Santisteban}, {Norberg}, {Peacock}, {Pier}, {Postman}, {Primak}, {Rae}, {Rai}, {Riess}, {Riffeser}, {Rix}, {R{\"o}ser}, {Russel}, {Rutz}, {Schilbach}, {Schultz}, {Scolnic}, {Strolger}, {Szalay}, {Seitz}, {Small}, {Smith}, {Soderblom}, {Taylor}, {Thomson}, {Taylor}, {Thakar}, {Thiel}, {Thilker}, {Unger}, {Urata}, {Valenti}, {Wagner}, {Walder}, {Walter}, {Watters}, {Werner}, {Wood-Vasey}, \& {Wyse}}]{Chambers+2016}
{Chambers}, K.~C., {Magnier}, E.~A., {Metcalfe}, N., {et~al.} 2016, arXiv e-prints, arXiv:1612.05560, \dodoi{10.48550/arXiv.1612.05560}

\bibitem[{{Chang} {et~al.}(2013){Chang}, {van der Wel}, {Rix}, {Holden}, {Bell}, {McGrath}, {Wuyts}, {H{\"a}ussler}, {Barden}, {Faber}, {Mozena}, {Ferguson}, {Guo}, {Galametz}, {Grogin}, {Kocevski}, {Koekemoer}, {Dekel}, {Huang}, {Hathi}, \& {Donley}}]{chang:2013b}
{Chang}, Y.-Y., {van der Wel}, A., {Rix}, H.-W., {et~al.} 2013, \apj, 773, 149, \dodoi{10.1088/0004-637X/773/2/149}

\bibitem[{{Clauwens} {et~al.}(2018){Clauwens}, {Schaye}, {Franx}, \& {Bower}}]{Clauwens:2018}
{Clauwens}, B., {Schaye}, J., {Franx}, M., \& {Bower}, R.~G. 2018, \mnras, 478, 3994, \dodoi{10.1093/mnras/sty1229}

\bibitem[{{Conroy} {et~al.}(2009){Conroy}, {Gunn}, \& {White}}]{Conroy:2009}
{Conroy}, C., {Gunn}, J.~E., \& {White}, M. 2009, \apj, 699, 486, \dodoi{10.1088/0004-637X/699/1/486}

\bibitem[{{Conselice}(2003)}]{Conselice:2003}
{Conselice}, C.~J. 2003, \apjs, 147, 1, \dodoi{10.1086/375001}

\bibitem[{{Conselice}(2014)}]{Conselice:2014}
---. 2014, \araa, 52, 291, \dodoi{10.1146/annurev-astro-081913-040037}

\bibitem[{{Crain} {et~al.}(2015){Crain}, {Schaye}, {Bower}, {Furlong}, {Schaller}, {Theuns}, {Dalla Vecchia}, {Frenk}, {McCarthy}, {Helly}, {Jenkins}, {Rosas-Guevara}, {White}, \& {Trayford}}]{Crain:2015}
{Crain}, R.~A., {Schaye}, J., {Bower}, R.~G., {et~al.} 2015, \mnras, 450, 1937, \dodoi{10.1093/mnras/stv725}

\bibitem[{{de Graaff} {et~al.}(2022){de Graaff}, {Trayford}, {Franx}, {Schaller}, {Schaye}, \& {van der Wel}}]{deGraaff:2022}
{de Graaff}, A., {Trayford}, J., {Franx}, M., {et~al.} 2022, \mnras, 511, 2544, \dodoi{10.1093/mnras/stab3510}

\bibitem[{{de Jong} {et~al.}(2013){de Jong}, {Verdoes Kleijn}, {Kuijken}, \& {Valentijn}}]{deJong:2013}
{de Jong}, J. T.~A., {Verdoes Kleijn}, G.~A., {Kuijken}, K.~H., \& {Valentijn}, E.~A. 2013, Experimental Astronomy, 35, 25, \dodoi{10.1007/s10686-012-9306-1}

\bibitem[{{de Vaucouleurs}(1959)}]{1959HDP+de_Vaucouleurs}
{de Vaucouleurs}, G. 1959, Handbuch der Physik, 53, 275, \dodoi{10.1007/978-3-642-45932-0_7}

\bibitem[{{Driver} {et~al.}(2009){Driver}, {Norberg}, {Baldry}, {Bamford}, {Hopkins}, {Liske}, {Loveday}, {Peacock}, {Hill}, {Kelvin}, {Robotham}, {Cross}, {Parkinson}, {Prescott}, {Conselice}, {Dunne}, {Brough}, {Jones}, {Sharp}, {van Kampen}, {Oliver}, {Roseboom}, {Bland-Hawthorn}, {Croom}, {Ellis}, {Cameron}, {Cole}, {Frenk}, {Couch}, {Graham}, {Proctor}, {De Propris}, {Doyle}, {Edmondson}, {Nichol}, {Thomas}, {Eales}, {Jarvis}, {Kuijken}, {Lahav}, {Madore}, {Seibert}, {Meyer}, {Staveley-Smith}, {Phillipps}, {Popescu}, {Sansom}, {Sutherland}, {Tuffs}, \& {Warren}}]{Driver:2009}
{Driver}, S.~P., {Norberg}, P., {Baldry}, I.~K., {et~al.} 2009, Astronomy and Geophysics, 50, 5.12, \dodoi{10.1111/j.1468-4004.2009.50512.x}

\bibitem[{{Driver} {et~al.}(2011){Driver}, {Hill}, {Kelvin}, {Robotham}, {Liske}, {Norberg}, {Baldry}, {Bamford}, {Hopkins}, {Loveday}, {Peacock}, {Andrae}, {Bland-Hawthorn}, {Brough}, {Brown}, {Cameron}, {Ching}, {Colless}, {Conselice}, {Croom}, {Cross}, {de Propris}, {Dye}, {Drinkwater}, {Ellis}, {Graham}, {Grootes}, {Gunawardhana}, {Jones}, {van Kampen}, {Maraston}, {Nichol}, {Parkinson}, {Phillipps}, {Pimbblet}, {Popescu}, {Prescott}, {Roseboom}, {Sadler}, {Sansom}, {Sharp}, {Smith}, {Taylor}, {Thomas}, {Tuffs}, {Wijesinghe}, {Dunne}, {Frenk}, {Jarvis}, {Madore}, {Meyer}, {Seibert}, {Staveley-Smith}, {Sutherland}, \& {Warren}}]{Driver:2011}
{Driver}, S.~P., {Hill}, D.~T., {Kelvin}, L.~S., {et~al.} 2011, \mnras, 413, 971, \dodoi{10.1111/j.1365-2966.2010.18188.x}

\bibitem[{{Dubois} {et~al.}(2014){Dubois}, {Pichon}, {Welker}, {Le Borgne}, {Devriendt}, {Laigle}, {Codis}, {Pogosyan}, {Arnouts}, {Benabed}, {Bertin}, {Blaizot}, {Bouchet}, {Cardoso}, {Colombi}, {de Lapparent}, {Desjacques}, {Gavazzi}, {Kassin}, {Kimm}, {McCracken}, {Milliard}, {Peirani}, {Prunet}, {Rouberol}, {Silk}, {Slyz}, {Sousbie}, {Teyssier}, {Tresse}, {Treyer}, {Vibert}, \& {Volonteri}}]{Dubois2014}
{Dubois}, Y., {Pichon}, C., {Welker}, C., {et~al.} 2014, \mnras, 444, 1453, \dodoi{10.1093/mnras/stu1227}

\bibitem[{{Eisert} {et~al.}(2024){Eisert}, {Bottrell}, {Pillepich}, {Shimakawa}, {Rodriguez-Gomez}, {Nelson}, {Angeloudi}, \& {Huertas-Company}}]{Eisert2024contrastive}
{Eisert}, L., {Bottrell}, C., {Pillepich}, A., {et~al.} 2024, \mnras, 528, 7411, \dodoi{10.1093/mnras/stae481}

\bibitem[{{Fouque} {et~al.}(1990){Fouque}, {Bottinelli}, {Gouguenheim}, \& {Paturel}}]{Fouque1990}
{Fouque}, P., {Bottinelli}, L., {Gouguenheim}, L., \& {Paturel}, G. 1990, \apj, 349, 1, \dodoi{10.1086/168288}

\bibitem[{{Gadotti}(2009)}]{Gadotti:2009}
{Gadotti}, D.~A. 2009, \mnras, 393, 1531, \dodoi{10.1111/j.1365-2966.2008.14257.x}

\bibitem[{{Gao} \& {Ho}(2017)}]{Gao+2017}
{Gao}, H., \& {Ho}, L.~C. 2017, \apj, 845, 114, \dodoi{10.3847/1538-4357/aa7da4}

\bibitem[{{Gao} {et~al.}(2018){Gao}, {Ho}, {Barth}, \& {Li}}]{Gao+2018}
{Gao}, H., {Ho}, L.~C., {Barth}, A.~J., \& {Li}, Z.-Y. 2018, \apj, 862, 100, \dodoi{10.3847/1538-4357/aacdac}

\bibitem[{{Gao} {et~al.}(2019){Gao}, {Ho}, {Barth}, \& {Li}}]{Gao:2019}
---. 2019, \apjs, 244, 34, \dodoi{10.3847/1538-4365/ab3c6a}

\bibitem[{{Gao} {et~al.}(2020){Gao}, {Ho}, {Barth}, \& {Li}}]{Gao+2020}
---. 2020, \apjs, 247, 20, \dodoi{10.3847/1538-4365/ab67b2}

\bibitem[{{Gao} {et~al.}(2022){Gao}, {Ho}, \& {Li}}]{Gao+2022}
{Gao}, H., {Ho}, L.~C., \& {Li}, Z.-Y. 2022, \apjs, 262, 54, \dodoi{10.3847/1538-4365/ac8dea}

\bibitem[{{Genel} {et~al.}(2015){Genel}, {Fall}, {Hernquist}, {Vogelsberger}, {Snyder}, {Rodriguez-Gomez}, {Sijacki}, \& {Springel}}]{Genel:2015}
{Genel}, S., {Fall}, S.~M., {Hernquist}, L., {et~al.} 2015, \apjl, 804, L40, \dodoi{10.1088/2041-8205/804/2/L40}

\bibitem[{{Guzm{\'a}n-Ortega} {et~al.}(2023){Guzm{\'a}n-Ortega}, {Rodriguez-Gomez}, {Snyder}, {Chamberlain}, \& {Hernquist}}]{Guzman-Ortega:2023}
{Guzm{\'a}n-Ortega}, A., {Rodriguez-Gomez}, V., {Snyder}, G.~F., {Chamberlain}, K., \& {Hernquist}, L. 2023, \mnras, 519, 4920, \dodoi{10.1093/mnras/stac3334}

\bibitem[{{Haslbauer} {et~al.}(2022){Haslbauer}, {Banik}, {Kroupa}, {Wittenburg}, \& {Javanmardi}}]{Haslbauer:2022}
{Haslbauer}, M., {Banik}, I., {Kroupa}, P., {Wittenburg}, N., \& {Javanmardi}, B. 2022, \apj, 925, 183, \dodoi{10.3847/1538-4357/ac46ac}

\bibitem[{{Hirschmann} {et~al.}(2014){Hirschmann}, {Dolag}, {Saro}, {Bachmann}, {Borgani}, \& {Burkert}}]{Hirschmann2014}
{Hirschmann}, M., {Dolag}, K., {Saro}, A., {et~al.} 2014, \mnras, 442, 2304, \dodoi{10.1093/mnras/stu1023}

\bibitem[{{Holden} {et~al.}(2012){Holden}, {van der Wel}, {Rix}, \& {Franx}}]{Holden:2012}
{Holden}, B.~P., {van der Wel}, A., {Rix}, H.-W., \& {Franx}, M. 2012, \apj, 749, 96, \dodoi{10.1088/0004-637X/749/2/96}

\bibitem[{{Huang} {et~al.}(2018){Huang}, {Leauthaud}, {Greene}, {Bundy}, {Lin}, {Tanaka}, {Miyazaki}, \& {Komiyama}}]{Huang:2018individualHalo}
{Huang}, S., {Leauthaud}, A., {Greene}, J.~E., {et~al.} 2018, \mnras, 475, 3348, \dodoi{10.1093/mnras/stx3200}

\bibitem[{{Hubble}(1926)}]{Hubble:1926}
{Hubble}, E.~P. 1926, \apj, 64, 321, \dodoi{10.1086/143018}

\bibitem[{{Hubble}(1936)}]{1936rene+Hubble}
---. 1936, {Realm of the Nebulae} (New Haven, CT: Yale Univ. Press)

\bibitem[{{Huertas-Company} {et~al.}(2019){Huertas-Company}, {Rodriguez-Gomez}, {Nelson}, {Pillepich}, {Bottrell}, {Bernardi}, {Dom{\'\i}nguez-S{\'a}nchez}, {Genel}, {Pakmor}, {Snyder}, \& {Vogelsberger}}]{Huertas-Company+2019}
{Huertas-Company}, M., {Rodriguez-Gomez}, V., {Nelson}, D., {et~al.} 2019, \mnras, 489, 1859, \dodoi{10.1093/mnras/stz2191}

\bibitem[{{Im} {et~al.}(1995){Im}, {Ratnatunga}, {Griffiths}, \& {Casertano}}]{Im:1995}
{Im}, M., {Ratnatunga}, K.~U., {Griffiths}, R.~E., \& {Casertano}, S. 1995, \apjl, 445, L15, \dodoi{10.1086/187878}

\bibitem[{{Kado-Fong} {et~al.}(2020{\natexlab{a}}){Kado-Fong}, {Greene}, {Greco}, {Beaton}, {Goulding}, {Johnson}, \& {Komiyama}}]{Kado-Fong:2020SFR}
{Kado-Fong}, E., {Greene}, J.~E., {Greco}, J.~P., {et~al.} 2020{\natexlab{a}}, \aj, 159, 103, \dodoi{10.3847/1538-3881/ab6ef3}

\bibitem[{{Kado-Fong} {et~al.}(2020{\natexlab{b}}){Kado-Fong}, {Greene}, {Huang}, {Beaton}, {Goulding}, \& {Komiyama}}]{Kado-Fong:2020Shape}
{Kado-Fong}, E., {Greene}, J.~E., {Huang}, S., {et~al.} 2020{\natexlab{b}}, \apj, 900, 163, \dodoi{10.3847/1538-4357/abacc2}

\bibitem[{{Kormendy} {et~al.}(2010){Kormendy}, {Drory}, {Bender}, \& {Cornell}}]{Kormendy:2010}
{Kormendy}, J., {Drory}, N., {Bender}, R., \& {Cornell}, M.~E. 2010, \apj, 723, 54, \dodoi{10.1088/0004-637X/723/1/54}

\bibitem[{{Kormendy} \& {Fisher}(2008)}]{2008ASPC+Kormendy}
{Kormendy}, J., \& {Fisher}, D.~B. 2008, in ASP Conf. Ser. 396, Formation and Evolution of Galaxy Disks, ed. J.~G. {Funes} \& E.~M. {Corsini} (San Francisco, CA: ASP), 297.
\newblock \doarXiv{0810.2534}

\bibitem[{{Kroupa}(2001)}]{Kroupa:2001}
{Kroupa}, P. 2001, \mnras, 322, 231, \dodoi{10.1046/j.1365-8711.2001.04022.x}

\bibitem[{{Lambas} {et~al.}(1992){Lambas}, {Maddox}, \& {Loveday}}]{Lambas:1992}
{Lambas}, D.~G., {Maddox}, S.~J., \& {Loveday}, J. 1992, \mnras, 258, 404, \dodoi{10.1093/mnras/258.2.404}

\bibitem[{{Law} {et~al.}(2012){Law}, {Steidel}, {Shapley}, {Nagy}, {Reddy}, \& {Erb}}]{Law:2012}
{Law}, D.~R., {Steidel}, C.~C., {Shapley}, A.~E., {et~al.} 2012, \apj, 745, 85, \dodoi{10.1088/0004-637X/745/1/85}

\bibitem[{{Li} {et~al.}(2022){Li}, {Huang}, {Leauthaud}, {Moustakas}, {Danieli}, {Greene}, {Abraham}, {Ardila}, {Kado-Fong}, {Lokhorst}, {Lupton}, \& {Price}}]{Li:2022}
{Li}, J., {Huang}, S., {Leauthaud}, A., {et~al.} 2022, \mnras, 515, 5335, \dodoi{10.1093/mnras/stac2121}

\bibitem[{{Liske} {et~al.}(2015){Liske}, {Baldry}, {Driver}, {Tuffs}, {Alpaslan}, {Andrae}, {Brough}, {Cluver}, {Grootes}, {Gunawardhana}, {Kelvin}, {Loveday}, {Robotham}, {Taylor}, {Bamford}, {Bland-Hawthorn}, {Brown}, {Drinkwater}, {Hopkins}, {Meyer}, {Norberg}, {Peacock}, {Agius}, {Andrews}, {Bauer}, {Ching}, {Colless}, {Conselice}, {Croom}, {Davies}, {De Propris}, {Dunne}, {Eardley}, {Ellis}, {Foster}, {Frenk}, {H{\"a}u{\ss}ler}, {Holwerda}, {Howlett}, {Ibarra}, {Jarvis}, {Jones}, {Kafle}, {Lacey}, {Lange}, {Lara-L{\'o}pez}, {L{\'o}pez-S{\'a}nchez}, {Maddox}, {Madore}, {McNaught-Roberts}, {Moffett}, {Nichol}, {Owers}, {Palamara}, {Penny}, {Phillipps}, {Pimbblet}, {Popescu}, {Prescott}, {Proctor}, {Sadler}, {Sansom}, {Seibert}, {Sharp}, {Sutherland}, {V{\'a}zquez-Mata}, {van Kampen}, {Wilkins}, {Williams}, \& {Wright}}]{Liske:2015}
{Liske}, J., {Baldry}, I.~K., {Driver}, S.~P., {et~al.} 2015, \mnras, 452, 2087, \dodoi{10.1093/mnras/stv1436}

\bibitem[{{Ludlow} {et~al.}(2021){Ludlow}, {Fall}, {Schaye}, \& {Obreschkow}}]{Ludlow:2021}
{Ludlow}, A.~D., {Fall}, S.~M., {Schaye}, J., \& {Obreschkow}, D. 2021, \mnras, 508, 5114, \dodoi{10.1093/mnras/stab2770}

\bibitem[{{Ludlow} {et~al.}(2023){Ludlow}, {Fall}, {Wilkinson}, {Schaye}, \& {Obreschkow}}]{Ludlow+2023}
{Ludlow}, A.~D., {Fall}, S.~M., {Wilkinson}, M.~J., {Schaye}, J., \& {Obreschkow}, D. 2023, \mnras, 525, 5614, \dodoi{10.1093/mnras/stad2615}

\bibitem[{{Masters} {et~al.}(2010){Masters}, {Nichol}, {Bamford}, {Mosleh}, {Lintott}, {Andreescu}, {Edmondson}, {Keel}, {Murray}, {Raddick}, {Schawinski}, {Slosar}, {Szalay}, {Thomas}, \& {Vandenberg}}]{Masters2010}
{Masters}, K.~L., {Nichol}, R., {Bamford}, S., {et~al.} 2010, \mnras, 404, 792, \dodoi{10.1111/j.1365-2966.2010.16335.x}

\bibitem[{{Merritt} {et~al.}(2016){Merritt}, {van Dokkum}, {Abraham}, \& {Zhang}}]{Merritt:2016}
{Merritt}, A., {van Dokkum}, P., {Abraham}, R., \& {Zhang}, J. 2016, \apj, 830, 62, \dodoi{10.3847/0004-637X/830/2/62}

\bibitem[{{Miyazaki} {et~al.}(2012){Miyazaki}, {Komiyama}, {Nakaya}, {Kamata}, {Doi}, {Hamana}, {Karoji}, {Furusawa}, {Kawanomoto}, {Morokuma}, {Ishizuka}, {Nariai}, {Tanaka}, {Uraguchi}, {Utsumi}, {Obuchi}, {Okura}, {Oguri}, {Takata}, {Tomono}, {Kurakami}, {Namikawa}, {Usuda}, {Yamanoi}, {Terai}, {Uekiyo}, {Yamada}, {Koike}, {Aihara}, {Fujimori}, {Mineo}, {Miyatake}, {Yasuda}, {Nishizawa}, {Saito}, {Tanaka}, {Uchida}, {Katayama}, {Wang}, {Chen}, {Lupton}, {Loomis}, {Bickerton}, {Price}, {Gunn}, {Suzuki}, {Miyazaki}, {Muramatsu}, {Yamamoto}, {Endo}, {Ezaki}, {Itoh}, {Miwa}, {Yokota}, {Matsuda}, {Ebinuma}, \& {Takeshi}}]{Miyazaki2012}
{Miyazaki}, S., {Komiyama}, Y., {Nakaya}, H., {et~al.} 2012, in Society of Photo-Optical Instrumentation Engineers (SPIE) Conference Series, Vol. 8446, Ground-based and Airborne Instrumentation for Astronomy IV, ed. I.~S. {McLean}, S.~K. {Ramsay}, \& H.~{Takami}, 84460Z, \dodoi{10.1117/12.926844}

\bibitem[{{Miyazaki} {et~al.}(2018){Miyazaki}, {Komiyama}, {Kawanomoto}, {Doi}, {Furusawa}, {Hamana}, {Hayashi}, {Ikeda}, {Kamata}, {Karoji}, {Koike}, {Kurakami}, {Miyama}, {Morokuma}, {Nakata}, {Namikawa}, {Nakaya}, {Nariai}, {Obuchi}, {Oishi}, {Okada}, {Okura}, {Tait}, {Takata}, {Tanaka}, {Tanaka}, {Terai}, {Tomono}, {Uraguchi}, {Usuda}, {Utsumi}, {Yamada}, {Yamanoi}, {Aihara}, {Fujimori}, {Mineo}, {Miyatake}, {Oguri}, {Uchida}, {Tanaka}, {Yasuda}, {Takada}, {Murayama}, {Nishizawa}, {Sugiyama}, {Chiba}, {Futamase}, {Wang}, {Chen}, {Ho}, {Liaw}, {Chiu}, {Ho}, {Lai}, {Lee}, {Jeng}, {Iwamura}, {Armstrong}, {Bickerton}, {Bosch}, {Gunn}, {Lupton}, {Loomis}, {Price}, {Smith}, {Strauss}, {Turner}, {Suzuki}, {Miyazaki}, {Muramatsu}, {Yamamoto}, {Endo}, {Ezaki}, {Ito}, {Kawaguchi}, {Sofuku}, {Taniike}, {Akutsu}, {Dojo}, {Kasumi}, {Matsuda}, {Imoto}, {Miwa}, {Suzuki}, {Takeshi}, \& {Yokota}}]{Miyazaki2018}
{Miyazaki}, S., {Komiyama}, Y., {Kawanomoto}, S., {et~al.} 2018, \pasj, 70, S1, \dodoi{10.1093/pasj/psx063}

\bibitem[{{Nelson} {et~al.}(2019{\natexlab{a}}){Nelson}, {Springel}, {Pillepich}, {Rodriguez-Gomez}, {Torrey}, {Genel}, {Vogelsberger}, {Pakmor}, {Marinacci}, {Weinberger}, {Kelley}, {Lovell}, {Diemer}, \& {Hernquist}}]{Nelson2019a}
{Nelson}, D., {Springel}, V., {Pillepich}, A., {et~al.} 2019{\natexlab{a}}, Computational Astrophysics and Cosmology, 6, 2, \dodoi{10.1186/s40668-019-0028-x}

\bibitem[{{Nelson} {et~al.}(2019{\natexlab{b}}){Nelson}, {Pillepich}, {Springel}, {Pakmor}, {Weinberger}, {Genel}, {Torrey}, {Vogelsberger}, {Marinacci}, \& {Hernquist}}]{Nelson2019outflow}
{Nelson}, D., {Pillepich}, A., {Springel}, V., {et~al.} 2019{\natexlab{b}}, \mnras, 490, 3234, \dodoi{10.1093/mnras/stz2306}

\bibitem[{{Padilla} \& {Strauss}(2008)}]{Padilla:2008}
{Padilla}, N.~D., \& {Strauss}, M.~A. 2008, \mnras, 388, 1321, \dodoi{10.1111/j.1365-2966.2008.13480.x}

\bibitem[{{Pandya} {et~al.}(2024){Pandya}, {Zhang}, {Huertas-Company}, {Iyer}, {McGrath}, {Barro}, {Finkelstein}, {K{\"u}mmel}, {Hartley}, {Ferguson}, {Kartaltepe}, {Primack}, {Dekel}, {Faber}, {Koo}, {Bryan}, {Somerville}, {Amor{\'\i}n}, {Arrabal Haro}, {Bagley}, {Bell}, {Bertin}, {Costantin}, {Dav{\'e}}, {Dickinson}, {Feldmann}, {Fontana}, {Gavazzi}, {Giavalisco}, {Grazian}, {Grogin}, {Guo}, {Hahn}, {Holwerda}, {Kewley}, {Kirkpatrick}, {Kocevski}, {Koekemoer}, {Lotz}, {Lucas}, {Papovich}, {Pentericci}, {P{\'e}rez-Gonz{\'a}lez}, {Pirzkal}, {Ravindranath}, {Rose}, {Schefer}, {Simons}, {Straughn}, {Tacchella}, {Trump}, {de la Vega}, {Wilkins}, {Wuyts}, {Yang}, \& {Yung}}]{Pandya+2024banana}
{Pandya}, V., {Zhang}, H., {Huertas-Company}, M., {et~al.} 2024, \apj, 963, 54, \dodoi{10.3847/1538-4357/ad1a13}

\bibitem[{{Pillepich} {et~al.}(2018){Pillepich}, {Springel}, {Nelson}, {Genel}, {Naiman}, {Pakmor}, {Hernquist}, {Torrey}, {Vogelsberger}, {Weinberger}, \& {Marinacci}}]{Pillepich:2018a}
{Pillepich}, A., {Springel}, V., {Nelson}, D., {et~al.} 2018, \mnras, 473, 4077, \dodoi{10.1093/mnras/stx2656}

\bibitem[{{Pillepich} {et~al.}(2019){Pillepich}, {Nelson}, {Springel}, {Pakmor}, {Torrey}, {Weinberger}, {Vogelsberger}, {Marinacci}, {Genel}, {van der Wel}, \& {Hernquist}}]{Pillepich2019}
{Pillepich}, A., {Nelson}, D., {Springel}, V., {et~al.} 2019, \mnras, 490, 3196, \dodoi{10.1093/mnras/stz2338}

\bibitem[{{Planck Collaboration} {et~al.}(2016){Planck Collaboration}, {Ade}, {Aghanim}, {Arnaud}, {Ashdown}, {Aumont}, {Baccigalupi}, {Banday}, {Barreiro}, {Bartlett}, {Bartolo}, {Battaner}, {Battye}, {Benabed}, {Beno{\^\i}t}, {Benoit-L{\'e}vy}, {Bernard}, {Bersanelli}, {Bielewicz}, {Bock}, {Bonaldi}, {Bonavera}, {Bond}, {Borrill}, {Bouchet}, {Boulanger}, {Bucher}, {Burigana}, {Butler}, {Calabrese}, {Cardoso}, {Catalano}, {Challinor}, {Chamballu}, {Chary}, {Chiang}, {Chluba}, {Christensen}, {Church}, {Clements}, {Colombi}, {Colombo}, {Combet}, {Coulais}, {Crill}, {Curto}, {Cuttaia}, {Danese}, {Davies}, {Davis}, {de Bernardis}, {de Rosa}, {de Zotti}, {Delabrouille}, {D{\'e}sert}, {Di Valentino}, {Dickinson}, {Diego}, {Dolag}, {Dole}, {Donzelli}, {Dor{\'e}}, {Douspis}, {Ducout}, {Dunkley}, {Dupac}, {Efstathiou}, {Elsner}, {En{\ss}lin}, {Eriksen}, {Farhang}, {Fergusson}, {Finelli}, {Forni}, {Frailis}, {Fraisse}, {Franceschi}, {Frejsel}, {Galeotta}, {Galli}, {Ganga}, {Gauthier}, {Gerbino}, {Ghosh}, {Giard}, {Giraud-H{\'e}raud}, {Giusarma}, {Gjerl{\o}w}, {Gonz{\'a}lez-Nuevo}, {G{\'o}rski}, {Gratton}, {Gregorio}, {Gruppuso}, {Gudmundsson}, {Hamann}, {Hansen}, {Hanson}, {Harrison}, {Helou}, {Henrot-Versill{\'e}}, {Hern{\'a}ndez-Monteagudo}, {Herranz}, {Hildebrandt}, {Hivon}, {Hobson}, {Holmes}, {Hornstrup}, {Hovest}, {Huang}, {Huffenberger}, {Hurier}, {Jaffe}, {Jaffe}, {Jones}, {Juvela}, {Keih{\"a}nen}, {Keskitalo}, {Kisner}, {Kneissl}, {Knoche}, {Knox}, {Kunz}, {Kurki-Suonio}, {Lagache}, {L{\"a}hteenm{\"a}ki}, {Lamarre}, {Lasenby}, {Lattanzi}, {Lawrence}, {Leahy}, {Leonardi}, {Lesgourgues}, {Levrier}, {Lewis}, {Liguori}, {Lilje}, {Linden-V{\o}rnle}, {L{\'o}pez-Caniego}, {Lubin}, {Mac{\'\i}as-P{\'e}rez}, {Maggio}, {Maino}, {Mandolesi}, {Mangilli}, {Marchini}, {Maris}, {Martin}, {Martinelli}, {Mart{\'\i}nez-Gonz{\'a}lez}, {Masi}, {Matarrese}, {McGehee}, {Meinhold}, {Melchiorri}, {Melin}, {Mendes}, {Mennella}, {Migliaccio}, {Millea}, {Mitra}, {Miville-Desch{\^e}nes}, {Moneti}, {Montier}, {Morgante}, {Mortlock}, {Moss}, {Munshi}, {Murphy}, {Naselsky}, {Nati}, {Natoli}, {Netterfield}, {N{\o}rgaard-Nielsen}, {Noviello}, {Novikov}, {Novikov}, {Oxborrow}, {Paci}, {Pagano}, {Pajot}, {Paladini}, {Paoletti}, {Partridge}, {Pasian}, {Patanchon}, {Pearson}, {Perdereau}, {Perotto}, {Perrotta}, {Pettorino}, {Piacentini}, {Piat}, {Pierpaoli}, {Pietrobon}, {Plaszczynski}, {Pointecouteau}, {Polenta}, {Popa}, {Pratt}, {Pr{\'e}zeau}, {Prunet}, {Puget}, {Rachen}, {Reach}, {Rebolo}, {Reinecke}, {Remazeilles}, {Renault}, {Renzi}, {Ristorcelli}, {Rocha}, {Rosset}, {Rossetti}, {Roudier}, {Rouill{\'e} d'Orfeuil}, {Rowan-Robinson}, {Rubi{\~n}o-Mart{\'\i}n}, {Rusholme}, {Said}, {Salvatelli}, {Salvati}, {Sandri}, {Santos}, {Savelainen}, {Savini}, {Scott}, {Seiffert}, {Serra}, {Shellard}, {Spencer}, {Spinelli}, {Stolyarov}, {Stompor}, {Sudiwala}, {Sunyaev}, {Sutton}, {Suur-Uski}, {Sygnet}, {Tauber}, {Terenzi}, {Toffolatti}, {Tomasi}, {Tristram}, {Trombetti}, {Tucci}, {Tuovinen}, {T{\"u}rler}, {Umana}, {Valenziano}, {Valiviita}, {Van Tent}, {Vielva}, {Villa}, {Wade}, {Wandelt}, {Wehus}, {White}, {White}, {Wilkinson}, {Yvon}, {Zacchei}, \& {Zonca}}]{PlanckCollaboration:2016}
{Planck Collaboration}, {Ade}, P.~A.~R., {Aghanim}, N., {et~al.} 2016, \aap, 594, A13, \dodoi{10.1051/0004-6361/201525830}

\bibitem[{{Popping} {et~al.}(2022){Popping}, {Pillepich}, {Calistro Rivera}, {Schulz}, {Hernquist}, {Kaasinen}, {Marinacci}, {Nelson}, \& {Vogelsberger}}]{Popping+2022}
{Popping}, G., {Pillepich}, A., {Calistro Rivera}, G., {et~al.} 2022, \mnras, 510, 3321, \dodoi{10.1093/mnras/stab3312}

\bibitem[{{R{\'e}my-Ruyer} {et~al.}(2014){R{\'e}my-Ruyer}, {Madden}, {Galliano}, {Galametz}, {Takeuchi}, {Asano}, {Zhukovska}, {Lebouteiller}, {Cormier}, {Jones}, {Bocchio}, {Baes}, {Bendo}, {Boquien}, {Boselli}, {DeLooze}, {Doublier-Pritchard}, {Hughes}, {Karczewski}, \& {Spinoglio}}]{Remy-Ruyer+2014}
{R{\'e}my-Ruyer}, A., {Madden}, S.~C., {Galliano}, F., {et~al.} 2014, \aap, 563, A31, \dodoi{10.1051/0004-6361/201322803}

\bibitem[{{Rix} {et~al.}(2004){Rix}, {Barden}, {Beckwith}, {Bell}, {Borch}, {Caldwell}, {H{\"a}ussler}, {Jahnke}, {Jogee}, {McIntosh}, {Meisenheimer}, {Peng}, {Sanchez}, {Somerville}, {Wisotzki}, \& {Wolf}}]{Rix:2004}
{Rix}, H.-W., {Barden}, M., {Beckwith}, S. V.~W., {et~al.} 2004, \apjs, 152, 163, \dodoi{10.1086/420885}

\bibitem[{{Rodr{\'\i}guez} \& {Padilla}(2013)}]{Rodriguez:2013}
{Rodr{\'\i}guez}, S., \& {Padilla}, N.~D. 2013, \mnras, 434, 2153, \dodoi{10.1093/mnras/stt1168}

\bibitem[{{Rodriguez-Gomez} {et~al.}(2019){Rodriguez-Gomez}, {Snyder}, {Lotz}, {Nelson}, {Pillepich}, {Springel}, {Genel}, {Weinberger}, {Tacchella}, {Pakmor}, {Torrey}, {Marinacci}, {Vogelsberger}, {Hernquist}, \& {Thilker}}]{Rodriguez-Gomez:2019}
{Rodriguez-Gomez}, V., {Snyder}, G.~F., {Lotz}, J.~M., {et~al.} 2019, \mnras, 483, 4140, \dodoi{10.1093/mnras/sty3345}

\bibitem[{{Ryden} \& {Terndrup}(1994)}]{Ryden:1994}
{Ryden}, B.~S., \& {Terndrup}, D.~M. 1994, \apj, 425, 43, \dodoi{10.1086/173960}

\bibitem[{{S{\'a}nchez-Janssen} {et~al.}(2010){S{\'a}nchez-Janssen}, {M{\'e}ndez-Abreu}, \& {Aguerri}}]{Sanchez-Janssen:2010}
{S{\'a}nchez-Janssen}, R., {M{\'e}ndez-Abreu}, J., \& {Aguerri}, J.~A.~L. 2010, \mnras, 406, L65, \dodoi{10.1111/j.1745-3933.2010.00883.x}

\bibitem[{{Sandage} {et~al.}(1970){Sandage}, {Freeman}, \& {Stokes}}]{Sandage:1970}
{Sandage}, A., {Freeman}, K.~C., \& {Stokes}, N.~R. 1970, \apj, 160, 831, \dodoi{10.1086/150475}

\bibitem[{{Schaye} {et~al.}(2015){Schaye}, {Crain}, {Bower}, {Furlong}, {Schaller}, {Theuns}, {Dalla Vecchia}, {Frenk}, {McCarthy}, {Helly}, {Jenkins}, {Rosas-Guevara}, {White}, {Baes}, {Booth}, {Camps}, {Navarro}, {Qu}, {Rahmati}, {Sawala}, {Thomas}, \& {Trayford}}]{Schaye:2015}
{Schaye}, J., {Crain}, R.~A., {Bower}, R.~G., {et~al.} 2015, \mnras, 446, 521, \dodoi{10.1093/mnras/stu2058}

\bibitem[{{S{\'e}rsic}(1968)}]{Sersic:1968}
{S{\'e}rsic}, J.~L. 1968, {Atlas de Galaxias Australes} (C{\'o}rdoba: Obs. Astron., Univ. Nac. C{\'o}rdoba)

\bibitem[{{Stone} {et~al.}(2021){Stone}, {Arora}, {Courteau}, \& {Cuillandre}}]{Stone:2021}
{Stone}, C.~J., {Arora}, N., {Courteau}, S., \& {Cuillandre}, J.-C. 2021, \mnras, 508, 1870, \dodoi{10.1093/mnras/stab2709}

\bibitem[{{Tacchella} {et~al.}(2019){Tacchella}, {Diemer}, {Hernquist}, {Genel}, {Marinacci}, {Nelson}, {Pillepich}, {Rodriguez-Gomez}, {Sales}, {Springel}, \& {Vogelsberger}}]{Tacchella:2019}
{Tacchella}, S., {Diemer}, B., {Hernquist}, L., {et~al.} 2019, \mnras, 487, 5416, \dodoi{10.1093/mnras/stz1657}

\bibitem[{{Taylor} {et~al.}(2011){Taylor}, {Hopkins}, {Baldry}, {Brown}, {Driver}, {Kelvin}, {Hill}, {Robotham}, {Bland-Hawthorn}, {Jones}, {Sharp}, {Thomas}, {Liske}, {Loveday}, {Norberg}, {Peacock}, {Bamford}, {Brough}, {Colless}, {Cameron}, {Conselice}, {Croom}, {Frenk}, {Gunawardhana}, {Kuijken}, {Nichol}, {Parkinson}, {Phillipps}, {Pimbblet}, {Popescu}, {Prescott}, {Sutherland}, {Tuffs}, {van Kampen}, \& {Wijesinghe}}]{Taylor:2011}
{Taylor}, E.~N., {Hopkins}, A.~M., {Baldry}, I.~K., {et~al.} 2011, \mnras, 418, 1587, \dodoi{10.1111/j.1365-2966.2011.19536.x}

\bibitem[{{Teklu} {et~al.}(2015){Teklu}, {Remus}, {Dolag}, {Beck}, {Burkert}, {Schmidt}, {Schulze}, \& {Steinborn}}]{Teklu2015}
{Teklu}, A.~F., {Remus}, R.-S., {Dolag}, K., {et~al.} 2015, \apj, 812, 29, \dodoi{10.1088/0004-637X/812/1/29}

\bibitem[{{Torrey} {et~al.}(2014){Torrey}, {Vogelsberger}, {Genel}, {Sijacki}, {Springel}, \& {Hernquist}}]{Torrey:2014}
{Torrey}, P., {Vogelsberger}, M., {Genel}, S., {et~al.} 2014, \mnras, 438, 1985, \dodoi{10.1093/mnras/stt2295}

\bibitem[{{van de Sande} {et~al.}(2019){van de Sande}, {Lagos}, {Welker}, {Bland-Hawthorn}, {Schulze}, {Remus}, {Bah{\'e}}, {Brough}, {Bryant}, {Cortese}, {Croom}, {Devriendt}, {Dubois}, {Goodwin}, {Konstantopoulos}, {Lawrence}, {Medling}, {Pichon}, {Richards}, {Sanchez}, {Scott}, \& {Sweet}}]{vandeSande:2019}
{van de Sande}, J., {Lagos}, C. D.~P., {Welker}, C., {et~al.} 2019, \mnras, 484, 869, \dodoi{10.1093/mnras/sty3506}

\bibitem[{{van der Wel} {et~al.}(2014){van der Wel}, {Chang}, {Bell}, {Holden}, {Ferguson}, {Giavalisco}, {Rix}, {Skelton}, {Whitaker}, {Momcheva}, {Brammer}, {Kassin}, {Martig}, {Dekel}, {Ceverino}, {Koo}, {Mozena}, {van Dokkum}, {Franx}, {Faber}, \& {Primack}}]{vanderWel:2014}
{van der Wel}, A., {Chang}, Y.-Y., {Bell}, E.~F., {et~al.} 2014, \apjl, 792, L6, \dodoi{10.1088/2041-8205/792/1/L6}

\bibitem[{{Vogelsberger} {et~al.}(2013){Vogelsberger}, {Genel}, {Sijacki}, {Torrey}, {Springel}, \& {Hernquist}}]{Vogelsberger:2013}
{Vogelsberger}, M., {Genel}, S., {Sijacki}, D., {et~al.} 2013, \mnras, 436, 3031, \dodoi{10.1093/mnras/stt1789}

\bibitem[{{Vogelsberger} {et~al.}(2014){Vogelsberger}, {Genel}, {Springel}, {Torrey}, {Sijacki}, {Xu}, {Snyder}, {Bird}, {Nelson}, \& {Hernquist}}]{Vogelsberger:2014}
{Vogelsberger}, M., {Genel}, S., {Springel}, V., {et~al.} 2014, \nat, 509, 177, \dodoi{10.1038/nature13316}

\bibitem[{{Weinberger} {et~al.}(2017){Weinberger}, {Springel}, {Hernquist}, {Pillepich}, {Marinacci}, {Pakmor}, {Nelson}, {Genel}, {Vogelsberger}, {Naiman}, \& {Torrey}}]{Weinberger:2017}
{Weinberger}, R., {Springel}, V., {Hernquist}, L., {et~al.} 2017, \mnras, 465, 3291, \dodoi{10.1093/mnras/stw2944}

\bibitem[{{Weinzirl} {et~al.}(2009){Weinzirl}, {Jogee}, {Khochfar}, {Burkert}, \& {Kormendy}}]{2009ApJ+Weinzirl}
{Weinzirl}, T., {Jogee}, S., {Khochfar}, S., {Burkert}, A., \& {Kormendy}, J. 2009, \apj, 696, 411, \dodoi{10.1088/0004-637X/696/1/411}

\bibitem[{{Wilkinson} {et~al.}(2023){Wilkinson}, {Ludlow}, {Lagos}, {Fall}, {Schaye}, \& {Obreschkow}}]{Wilkinson:2023}
{Wilkinson}, M.~J., {Ludlow}, A.~D., {Lagos}, C. d.~P., {et~al.} 2023, \mnras, 519, 5942, \dodoi{10.1093/mnras/stad055}

\bibitem[{{York} {et~al.}(2000){York}, {Adelman}, {Anderson}, {Anderson}, {Annis}, {Bahcall}, {Bakken}, {Barkhouser}, {Bastian}, {Berman}, {Boroski}, {Bracker}, {Briegel}, {Briggs}, {Brinkmann}, {Brunner}, {Burles}, {Carey}, {Carr}, {Castander}, {Chen}, {Colestock}, {Connolly}, {Crocker}, {Csabai}, {Czarapata}, {Davis}, {Doi}, {Dombeck}, {Eisenstein}, {Ellman}, {Elms}, {Evans}, {Fan}, {Federwitz}, {Fiscelli}, {Friedman}, {Frieman}, {Fukugita}, {Gillespie}, {Gunn}, {Gurbani}, {de Haas}, {Haldeman}, {Harris}, {Hayes}, {Heckman}, {Hennessy}, {Hindsley}, {Holm}, {Holmgren}, {Huang}, {Hull}, {Husby}, {Ichikawa}, {Ichikawa}, {Ivezi{\'c}}, {Kent}, {Kim}, {Kinney}, {Klaene}, {Kleinman}, {Kleinman}, {Knapp}, {Korienek}, {Kron}, {Kunszt}, {Lamb}, {Lee}, {Leger}, {Limmongkol}, {Lindenmeyer}, {Long}, {Loomis}, {Loveday}, {Lucinio}, {Lupton}, {MacKinnon}, {Mannery}, {Mantsch}, {Margon}, {McGehee}, {McKay}, {Meiksin}, {Merelli}, {Monet}, {Munn}, {Narayanan}, {Nash}, {Neilsen}, {Neswold}, {Newberg}, {Nichol}, {Nicinski}, {Nonino}, {Okada}, {Okamura}, {Ostriker}, {Owen}, {Pauls}, {Peoples}, {Peterson}, {Petravick}, {Pier}, {Pope}, {Pordes}, {Prosapio}, {Rechenmacher}, {Quinn}, {Richards}, {Richmond}, {Rivetta}, {Rockosi}, {Ruthmansdorfer}, {Sandford}, {Schlegel}, {Schneider}, {Sekiguchi}, {Sergey}, {Shimasaku}, {Siegmund}, {Smee}, {Smith}, {Snedden}, {Stone}, {Stoughton}, {Strauss}, {Stubbs}, {SubbaRao}, {Szalay}, {Szapudi}, {Szokoly}, {Thakar}, {Tremonti}, {Tucker}, {Uomoto}, {Vanden Berk}, {Vogeley}, {Waddell}, {Wang}, {Watanabe}, {Weinberg}, {Yanny}, {Yasuda}, \& {SDSS Collaboration}}]{SDSSCollaboration:2000}
{York}, D.~G., {Adelman}, J., {Anderson}, John~E., J., {et~al.} 2000, \aj, 120, 1579, \dodoi{10.1086/301513}

\bibitem[{{Zhang} {et~al.}(2022){Zhang}, {Wuyts}, {Witten}, {Avery}, {Hao}, {Sharma}, {Shen}, {Toshikawa}, \& {Villforth}}]{Zhang:2022threeDintrin}
{Zhang}, J., {Wuyts}, S., {Witten}, C., {et~al.} 2022, \mnras, 513, 4814, \dodoi{10.1093/mnras/stac1083}

\end{thebibliography}
\bibliographystyle{aasjournal}

\end{CJK*}
\end{document}